\documentclass[10pt]{article}
\usepackage{graphicx} 
\usepackage[a4paper, total={7in, 9in}]{geometry}
\usepackage{amsmath}
\usepackage[OT1]{fontenc}
\usepackage{xcolor}
\usepackage{algorithm}
\usepackage{algpseudocode}
\usepackage{hyperref}
\usepackage{booktabs}
\hypersetup{
  breaklinks=true,
  colorlinks=true,
  citecolor=blue,
  urlcolor=blue,
  linkcolor=blue
}
\usepackage[sorting=none,style=trad-unsrt]{biblatex}
\usepackage{authblk}
\usepackage{caption}
\DeclareCaptionLabelFormat{sfig}{#2 Fig}
\DeclareCaptionLabelFormat{stable}{#2 Table}
\AtEveryBibitem{%

}
\AtEveryBibitem{
      \ifboolexpr{ test {\ifentrytype{software}} or test {\ifentrytype{online}} }
    {}
    {\clearfield{url}\clearfield{urldate}}%
    \clearfield{urlyear}
    \clearfield{urlmonth}
    \clearfield{month}
    \clearfield{day}
    \clearname{editor}
    \clearfield{edition}
    \clearfield{issn}
    \clearfield{isbn}
    \clearname{translator}
    \clearfield{series}%
  \clearlist{location}%
  \clearfield{pagetotal}
  \clearfield{version}  \clearfield{note}
    \clearfield{booktitle}
}
\DeclareFieldFormat{doi}{\space\url{https://doi.org/#1}}
\DeclareFieldFormat[article]{title}{\MakeSentenceCase*{#1}}
\title{Socio-cognitive models in a patch foraging setting:\\ a case study for model selection and parameter identifiability methods}

\author[1]{Lisa Blum Moyse}
\author[1,2]{Ahmed El Hady}

\affil[1]{Centre for the Advanced Study of Collective Behavior, Konstanz, University of Konstanz, Konstanz, Germany }
\affil[2]{Department of Collective Behavior, Max Planck Institute of Animal Behavior, Konstanz, Germany}

\date{}

\begin{document}

\maketitle

\begin{abstract}
The mechanisms by which organisms extract relevant information from complex sensory signals and use it for decision-making constitute a fundamental issue in biological and cognitive sciences. In social animal societies, individual decision-making is profoundly shaped by information provided by conspecifics. Collective patch-foraging experiments in the laboratory provide a controlled setting in which social information use can be quantified. Here, we consider a go/no-go task in which groups choose between two patches differing in food reward probability. We use agent-based simulations, underpinned by an augmented collective drift-diffusion model, to investigate alternative mechanisms for representing and integrating social information. We consider two representations, continuous (\textit{counting representation}) and discrete (\textit{pulsatile representation}), and two integration mechanisms, biasing the decision threshold (\textit{threshold modulation}) and modifying the accumulated belief (\textit{belief modulation}), yielding four distinct social models, in addition to a non-interacting model. We first characterize the collective dynamics generated by these models across cognitive parameters and experimental conditions. The temporal dynamics of group accuracy provide informative signatures of the underlying mechanisms. We then assess model selection and parameter identifiability using Bayesian inference and Wasserstein distance minimization, applied to group-accuracy and departure-time distributions. Bayesian inference using distributions of group accuracy provides the most reliable identification across the conditions considered. Importantly, model selection and parameter identifiability are associated with different aspects of collective behavior: model selection is closely linked to the presence of oscillations in the temporal dynamics, whereas parameter identifiability is more closely related to the global accuracy value. Thus, the information available for distinguishing the underlying cognitive mechanisms is not necessarily the same as that required to recover their parameters. Our results further show that identification depends not only on model structure but also on the experimental conditions. Overall, this study provides practical guidelines for identifying socio-cognitive mechanisms in collective foraging experiments.
\end{abstract}

\section{Introduction}
How organisms extract relevant information from complex sensory inputs and use it to guide their decisions remains a key question in cognitive science~\cite{goldNeuralBasisDecision2007} and behavioral ecology~\cite{tinbergenStudyInstinct1989}. The same external stimulus can be represented differently depending on an organism's sensory and cognitive capabilities, a concept formalized by the notion of ``Umwelt'', the species-specific way in which organisms perceive and experience their environment~\cite{uexkullForayWorldsAnimals2010}.

\begin{figure}[!h]
    \centering
\includegraphics[width=0.8\linewidth]{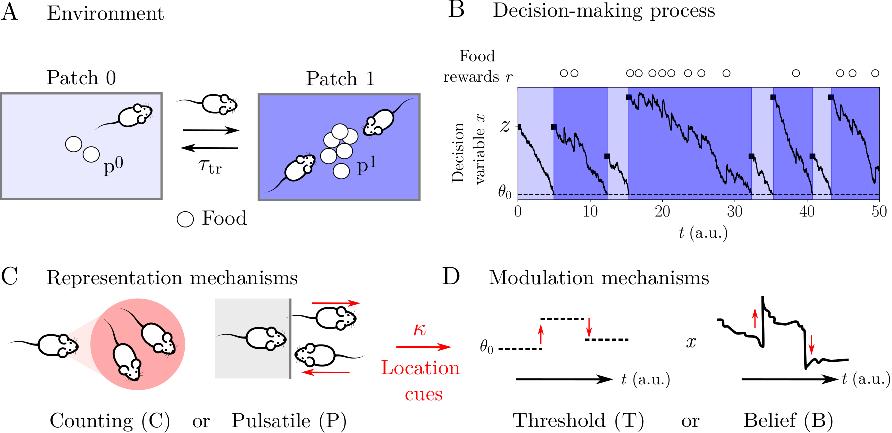}
    \caption{{\textbf{Experimental setup and cognitive models.}} (A) Two-patch foraging environment where animals decide to stay or move based on accumulated evidence. Patches differ in food abundance, with patch 1 being more rewarding than patch 0. Travel between patches lasts $\tau_\text{tr}$. (B) Decision mechanism: evidence accumulates in a decision variable $x$, starting from an initial bias $z$, until reaching a threshold $\theta_0$, which triggers a switch to the alternative patch. Within a patch, individuals accumulate both personal information (food rewards, $r$) and social information from observing conspecifics. (C) Two representations of social information: \textit{Counting (C)}, a continuous estimate of local group density or proportion, and \textit{Pulsatile (P)}, a discrete representation based on observed arrivals and departures. (D) Two modulation modes by which social information influences decisions: \textit{Threshold (T)}, which adjusts the decision threshold $\theta_0$ to change the propensity to switch patches, and \textit{Belief (B)}, which contributes social observations directly to the evidence accumulation process, modifying $x$.}
    \label{fig:schema}
\end{figure}

To investigate rigorously these processes, a powerful approach is to examine ecologically relevant decision processes, while preserving simple environmental structures; so that conditions can be precisely controlled and linked to the resulting behavioral observables. With that in mind, foraging, a fundamental behavior observed in most animal species~\cite{stephensForagingBehaviorEcology2007a}, monitored in laboratory settings, provides a notably useful framework.
Among the possible experiments, those with food patches constitute a rigorous substrate for go/no-go decision tasks: individuals accumulate information about the quality of the food patch, which influences their decision to stay or leave to explore another patch~\cite{charnovOptimalForagingMarginal1976,haydenNeuronalBasisSequential2011,constantinoLearningOpportunityCost2015,harhenOverharvestingHumanPatch2023,lloydUnderstandingPatchForaging2023}. In this paper, we study the minimal case of a two-patch foraging environment, see Fig~\ref{fig:schema}A; a conceptually simple framework, yet yielding rich behavioral dynamics.

To gain insight into this system and link observable behavior to internal decision dynamics, computational models are necessary tools. In particular, the drift-diffusion model~\cite{ratcliffDiffusionDecisionModel2008} provides an appropriate framework: decisions arise from the accumulation of noisy evidence over time until a decision threshold is reached. This model has established neural correlates~\cite{liuNeuralCorrelatesEvidence2011,hukNeuralCorrelatesNeural2012} and has illuminated the neural basis of binary perceptual choices~\cite{goldNeuralBasisDecision2007}. Importantly, this model naturally incorporates uncertainty and noise in decision processes-factors, that strongly shape behavior but are often neglected in ecological models. The framework has been adapted to patch-foraging decisions~\cite{davidsonForagingEvidenceAccumulation2019a,kilpatrickUncertaintyDrivesDeviations2021a}, particularly for two-patch settings~\cite{blummoyseSocialPatchForaging2025a,blummoyseSocialHierarchyShapes2026}, see Fig~\ref{fig:schema}B. Within this framework, different information sources can influence decisions in different ways: they may contribute directly to evidence accumulation or bias the decision process by modulating the decision threshold.

These information sources available to an individual can be personal, like direct observation of the physical environment, but can also be social. Indeed, many species forage in groups, where conspecifics become part of the sensory landscape, providing then cues that influence how individuals perceive and respond to their surroundings. Agents can observe intentional or inadvertent signals, such as the presence of a conspecific in a patch. This second type of signaling, which constitutes the focus of this study, is named \emph{location cues}. It is widespread across taxa; some examples include birds preferentially foraging where conspecifics are feeding~\cite{waiteCopyingForagingLocations1988}, or Norway rats following conspecifics from burrows to food sources~\cite{calhounEcologySociologyNorway1963}. The location cues reveal the conclusions the observed individuals have drawn~\cite{dallInformationItsUse2005,galefSocialInfluencesForaging2001,leadbeaterSocialLearningInsects2007,lihoreauLocalEnhancementPromotes2011}, thus increasing what is known about the environment. 

Combining this collective aspect with the questions related to decision-making cognition then raises a fundamental issue: \emph{how is social information represented and how does it enter the decision process?}  To address this problem, we conceptualized two possible internal representations~\cite{blummoyseSocialPatchForaging2025a,blummoyseSocialHierarchyShapes2026} (i) and two possible integration mechanisms (ii) of location cues. 

(i) Individuals can continuously monitor the number of conspecifics at different locations, effectively estimating the proportion of the group occupying each patch: a \textit{counting representation (C)} based on continuous social density estimates, see Fig~\ref{fig:schema}C. Alternatively, social information can be represented as discrete behavioral events, such as observing the arrival or departure~\cite{bidariStochasticDynamicsSocial2022a} of individuals to/from a patch. In this \textit{pulsatile representation (P)}, information is transmitted only when behavioral transitions occur (e.g., patch switches), rather than through continuous monitoring.

(ii) Beyond how social information is represented, a critical question concerns how it is integrated into individual decision-making. Within the evidence accumulation framework, we consider two mechanisms by which social information can influence decisions, see Fig~\ref{fig:schema}D. First, social information can modulate the decision threshold (a principle already investigated with collapsing bounds driven by urgency in perceptual decision-making~\cite{smithModelingEvidenceAccumulation2022}). We name this mechanism \textit{threshold modulation (T)}. For example, a higher density of conspecifics in a patch may increase the tendency to remain by reducing the threshold for continued exploitation. Alternatively, social information can contribute directly to the evidence stream: this is \textit{belief modulation (B)}. Observations of conspecifics are then treated as additional evidence on patch quality and dynamically integrated alongside personal information~\cite{ratcliffComparisonSequentialSampling2004}. In this case, social observations accumulate like any other source of evidence, shaping both the speed and direction of the decision process.

Alongside a model with non-interacting agents (NI), combining these dimensions; (i) two representations (counting or pulsatile representation) and (ii) two integration modes (threshold or belief modulation), yields four socio-cognitive models of patch foraging: counting representation with threshold modulation (CT), counting representation with belief modulation (CB), pulsatile representation with Threshold modulation (PT), and pulsatile representation with belief modulation (PB).

Although these models rely on distinct underlying computations, they can generate qualitatively similar macroscopic outcomes, such as aggregation in high-quality patches. This raises a crucial challenge, which should be addressed before the application of models to empirical data~\cite{lillacciParameterEstimationModel2010,clermontInverseProblemMathematical2015,heathcoteIntroductionGoodPractices2015,villaverdeStructuralIdentifiabilityDynamic2016,wilsonTenSimpleRules2019,lukenParameterIdentifiabilityEvidenceaccumulation2025}: \emph{can different cognitive mechanisms be distinguished from collective behavioral observations?} Addressing this question requires assessing the selection of the models and the identifiability of their parameters; i.e., determining whether distinct mechanisms can be uniquely recovered from empirical data. 
Please note that in this article we will use the umbrella term \emph{identification} when referring to both model selection and parameter identifiability.

In that regard, in this study we developed a framework to evaluate the identification of the different cognitive models, which could then guide the design of collective decision-making experiments. We assessed the efficiency of two methods for estimating models and parameters: Bayesian inference with maximum a posteriori estimation~\cite{gelmanBayesianDataAnalysis2013} and Wasserstein distance minimization~\cite{berntonParameterEstimationWasserstein2019}, based either on distributions sampled at different time points of group accuracy; defined by the fraction of agents in the best patch, or on the distribution of departure times. A primary characterization of the computational models allowed us to gain useful insights on the conditions of identification.

We showed that the temporal dynamics of group accuracy provides informative signatures of the underlying mechanisms, with patterns of damped oscillations. We quantified them throughout the article with three metrics reflecting residence time, the global value of group accuracy, and the rate of change (larger with oscillations). We found that the identification of the collective models depends both on the richness of the collective behavioral signature and the way behavioral data are summarized. Bayesian inference applied to the distribution of group accuracy provided the most reliable identification. Interestingly, across models and experimental conditions, model selection success appeared to be closely associated with the group accuracy rate of change (showing up as oscillations), while parameter identifiability was more correlated with global accuracy. 

More broadly, this work provides a general framework for linking observable collective dynamics to the latent dynamical mechanisms that generate them. By systematically separating how social information is represented from how that information influences decisions, our approach contributes to understanding how organisms extract, encode, and use information from their social environment. Our long-term aim with this study is to provide general guidelines and methods for applying the models to observations.

\section{Methods}

Our code in Python is publicly available at~\cite{LblummoyseTwopatchforagingidentifiability}. 

All parameters, variables, models, collective behavior metrics, and identification measurements are summarized in Table~\ref{tab:notation}.

\subsection{Computational model}\label{methods:behav}

\subsubsection{Environment}

Agents forage between two patches ($k = 0$ and $k = 1$) separated by travel time $\tau_\text{tr}$. Within a patch, an agent receives a food reward with probability $p^k$ (where $p^0 \leq p^1$), followed by a refractory period $\tau_\text{rw}$ before the next reward opportunity. 

\textbf{Food depletion:} Resources can remain constant or deplete when consumed. For depletion dynamics, the reward probability is $p^k = A^k / A_m$, where $A^k$ is the current food amount in patch $k$ and $A_m$ is the maximum. Each consumed reward decrements $A^k \leftarrow A^k - 1$. When $A^k < p^k A_m$, the patch replenishes according to $A^k \leftarrow a$ at each time step, where $a$ is the replenishment factor.

\subsubsection{Non-interacting decision-making model (NI)}

The decision to leave a patch is governed by a drift-diffusion model. For agent $i$ in patch $k$, the decision variable $x_i$ evolves as:
\begin{equation}
dx_i = (r_i^k - \alpha_i) \, dt + \sigma_i \, dW_i
\label{eq:xinc}
\end{equation}
with initial condition $x_i = 0$. When $x_i$ reaches threshold $\theta_{0,i}$ (with $\theta_{0,i} < 0$), the agent switches to the alternative patch and $x_i$ resets to $z_i^k$. Here, $r_i^k$ is the food reward (1 if reward obtained, 0 otherwise), $\alpha_i$ is the drift (foraging cost), $\sigma_i$ is the noise amplitude, and $W_i(t)$ is a standard Wiener process.

\textbf{Initial bias - learning:} To represent learned estimates of patch quality, the initial bias $z_i^k$ updates each time the agent switches patches:
\begin{equation}
z_i^k = \zeta_i \left( \sum_l \frac{r^k_{i,l}}{T^k_{i,l}} - \sum_{l'} \frac{r^{1-k}_{i,l'}}{T^{1-k}_{i,l'}} \right)
\label{eq:zbias}
\end{equation}
where $T^k_{i,l}$ is the residence time in patch $k$ during visit $l$, and $\zeta_i$ is the bias strength. A higher reward rate in patch $k$ produces a positive bias in that patch and a negative bias in the other, respectively reducing and increasing departure likelihood. At $t = 0$, $z_i^k = 0$.

\subsubsection{Social information representation and modulation of decisions}

The basic NI model is extended to include social location information through a coupling term $J^k$, whose effect depends on the representation and modulation mechanism.

\paragraph{Representation mechanisms:}

Social information is encoded as $J^k$, which quantifies the local social environment in patch $k$.

\textit{Counting representation (C):} A continuous measure of local group density,
\begin{equation}
J^k = \frac{1}{N-1} \left( -1 + \sum_{j \text{ in patch } k} 1 \right) - \frac{1}{2}
\label{eq:Jcount}
\end{equation}
where $N$ is group size. The first term is the fraction of other agents in patch $k$ (ranging from 0 to 1). Positive $J^k$ indicates high local density; negative $J^k$ indicates low density.

\textit{Pulsatile representation (P):} A discrete measure based on arrivals and departures (we show here the time dependency to highlight the travel time $\tau_\mathrm{tr}$ shift),
\begin{equation}
J^k(t) = \frac{1}{N-1} \left( -\sum_{j \text{ in patch } k} \delta\bigl(x_j(t) - \theta_j^k(t)\bigr) + \sum_{j \text{ in patch } 1-k} \delta\bigl(x_j(t-\tau_\mathrm{tr}) - \theta_j^{1-k}(t-\tau_\mathrm{tr})\bigr) \right)
\label{eq:Jpuls}
\end{equation}
The first (resp. second) term is the sum of departures from (resp. arrivals to) patch $k$, normalized by group size minus one, the observer. Negative $J^k$ reflects more departures than arrivals; positive $J^k$ reflects the opposite.

\paragraph{Modulation mechanisms:}

These representations influence decisions through two distinct mechanisms:

\textit{Threshold modulation (T):} Social information biases the propensity to leave or stay without modifying the internal evidence accumulation. This is implemented by modulating the decision threshold:
\begin{equation}
\theta_i^k = \theta_{0,i} \left( 1 + \kappa_i J^k \right)
\label{eq:theta_mod}
\end{equation}
where $\kappa_i$ is the coupling strength. Eq~\eqref{eq:xinc} is unchanged. We remain in the modulation regime of $\kappa_i |J_k|<1$. When $J^k < 0$ (low density or many departures), $\theta_i^k$ decreases in magnitude (less negative), increasing the bias towards leaving. When $J^k > 0$, the threshold becomes more negative, decreasing the bias towards leaving.

To maintain physical meaning, we impose that the threshold is always lower than the initial bias ($\theta_i^k<z_i^k$). Given the parameter space ranges specified below, this constraint corresponds to $\theta_i^k \in [-8.25,-1.75]$ and $z_i^k\in[-1.5,1.5]$.

\textit{Belief modulation (B):} Social information contributes directly to evidence accumulation. For the counting representation:
\begin{equation}
dx_i = (r_i^k - \alpha_i + \kappa_i J^k) \, dt + \sigma_i \, dW_i
\label{eq:belief_count}
\end{equation}
For the pulsatile representation, as it is a jump integrated into the continuous dynamics~\cite{gardinerHandbookStochasticMethods1985a}:
\begin{equation}
dx_i = (r_i^k - \alpha_i) \, dt + \kappa_i J^k + \sigma_i \, dW_i
\label{eq:belief_puls}
\end{equation}
The threshold remains constant. As before, negative $J^k$ shifts evidence toward leaving; positive $J^k$ shifts evidence toward staying.

\subsubsection{Individual heterogeneity}

To model variation among agents, each parameter $u_i \in \{\alpha_i, \theta_{0,i}, \sigma_i, \zeta_i, \kappa_i\}$ is drawn from a normal distribution with mean $\overline{u}$ and standard deviation $\beta \cdot \overline{u}$, where $\beta$ is the heterogeneity factor.

\begin{table}[!h]
    \centering
    \begin{tabular}{|l|p{12cm}|}
    \hline
         \textbf{Symbol} & \textbf{Definition}\\
         \hline
         \multicolumn{2}{|c|}{\textit{Experimental parameters}} \\
         \hline
         $k$ & Patch index (0 or 1)\\
         $p^k$ & Reward probability in patch $k$\\
         $A^k$ & Current food amount in patch $k$\\
         $A_m$ & Maximum food amount per patch\\
         $a$ & Replenishment factor\\
         $\tau_\text{tr}$ & Travel time between patches\\
         $\tau_\text{rw}$ & Refractory period after reward\\
         $Q_\mathrm{in}$ & Initial group accuracy\\
         $N$ & Group size\\
         \hline
         \multicolumn{2}{|c|}{\textit{Decision model parameters}} \\
         \hline
         $\overline{\theta_0}$ & Decision threshold (mean)\\
         $\overline{\alpha}$ & Drift / foraging cost (mean)\\
         $\overline{\sigma}$ & Diffusion coefficient (mean)\\
         $\overline{\zeta}$ & Initial bias strength (mean)\\
         $\overline{\kappa}$ & Social coupling strength (mean)\\
         $\beta$ & Heterogeneity factor (scales inter-individual variations)\\
         \hline
         \multicolumn{2}{|c|}{\textit{Decision model variables}} \\
         \hline
         $x_i$ & Decision variable (accumulated evidence) for agent $i$\\
         $z_i^k$ & Initial bias for agent $i$ in patch $k$\\
         $r_i^k$ & Reward signal for agent $i$ in patch $k$\\
         $W_i$ & Wiener process (noise) for agent $i$\\
         $J^k$ & Social coupling term in patch $k$\\
         \hline
         \multicolumn{2}{|c|}{\textit{Representation and modulation mechanisms}} \\
         \hline
         {C} & Counting representation (continuous density estimate)\\
         {P} & Pulsatile representation (discrete arrivals/departures)\\
         {T} & Threshold modulation (adjusts decision threshold)\\
         {B} & Belief modulation (contributes to evidence)\\
         \hline
         \multicolumn{2}{|c|}{\textit{Models}} \\
         \hline
         NI & Non-interacting (no social information)\\
         CT & Counting + Threshold \\
         CB & Counting + Belief \\
         PT & Pulsatile + Threshold \\
         PB & Pulsatile + Belief \\
         \hline
         \multicolumn{2}{|c|}{\textit{Collective behavior metrics}} \\
         \hline
         $q$ & Group accuracy/ fraction of agents in the best patch (as a function of time)\\
         $\overline{q}$ & Mean accuracy over all groups (as a function of time)\\
         $T_{res}$ & Grand mean residence time in the best patch\\
         $Q$ & Grand mean accuracy\\
         $D$ & Mean rate of change of $\overline{q}$\\
         $D_\mathrm{group}$ & Mean rate of change of $q$\\
         \hline
         \multicolumn{2}{|c|}{\textit{Identification measurement}} \\
         \hline
         $\mathcal{M}$ & Proportion of correct model prediction\\
         $\mathcal{P}$ & Proportion of correct parameter prediction\\
         \hline
    \end{tabular}
    \caption{\textbf{Notation.} Definitions of parameters, variables, models, collective behavior metrics, and identification measurements. Overbars denote population means; subscripts $i$ denote individual values, superscripts $k$ denote the patch number.}
    \label{tab:notation}
\end{table}

\subsection{Simulation settings and numerical methods}

\subsubsection{Parameters}

\paragraph{Degeneracy resolution:} Drift-diffusion models exhibit parameter degeneracy~\cite{coxTheoryStochasticProcesses1965}, where multiple parameter combinations produce equivalent dynamics. To resolve this, one of $\{\overline{\theta_0}, \overline{\alpha}, \overline{\sigma}\}$ must be fixed~\cite{vandekerckhoveFittingRatcliffDiffusion2007}. Without loss of generality, we set $\overline{\sigma} = 0.1 \, \text{a.u.}^{-1}$.

\paragraph{Scaling coupling strength:} Pulsatile representation produces discrete jumps that require larger coupling strength than the continuous counting representation~\cite{blummoyseSocialPatchForaging2025a}. For visual clarity, we scaled the pulsatile coupling as $\overline{\kappa}_\mathrm{PT} = 3\overline{\kappa}$ and $\overline{\kappa}_\mathrm{PB} = 4\overline{\kappa}$.

\paragraph{Unit convention:}
For brevity, we omit explicit units for $\overline{\kappa}$ and $\beta$ in subsequent sections, noting that: (i) $\overline{\kappa}$ is dimensionless for models CT, PT, and PB, but has units a.u.${^{-1}}$ for model CB; (ii) Units of $\beta$ depend on the mean parameter so that the standard deviation is dimensionless.

\paragraph{Parameter space:} The parameter space spans:
$\overline{\theta_0} \in [-6, -4], \
\overline{\alpha} \in [0.75, 1.2]\, \text{a.u.}^{-1}, \
\overline{\zeta} \in [0, 6], \
\overline{\kappa} \in [0.3, 0.9],\
\beta \in [0, 0.15]$.
These boundaries were chosen to avoid saturation from excessively strong parameter values while maintaining biological plausibility.

The ranges for $\overline{\theta_0}$ and $\overline{\alpha}$ are constrained by the assumption that residence times remain within $T_\text{res} \in [3, 8] \, \text{a.u.}$, using the approximation $T_\text{res} \approx -\overline{\theta_0}/\overline{\alpha}$~\cite{blummoyseSocialPatchForaging2025a}. In such regime agents spend sufficient time at each patch to observe social information, yet switch frequently enough to explore both options.

We sampled the parameter space at values $n_l = 3$ per parameter, resulting in $3^5 = 243$ dynamics per model and condition. We chose $n_l = 3$ to balance computational cost and resolution; robustness to coarser and finer grids is documented in \ref{fig:bignl}~Fig.

\subsubsection{Numerical scheme} The decision variable Eq~\eqref{eq:xinc} is integrated using the Euler method with time step $dt = 0.1 \, \text{a.u.}$ All other variables are updated at each time step.

\subsubsection{Dynamics characterization} Fig~\ref{fig:dyn1} shows group accuracy (fraction of agents in the best patch) distributions $q$ at a given time (Fig~\ref{fig:dyn1}B), and its average across all groups $\overline{q}$ as a function of time. This example was simulated with the following parameter settings: $\overline{\alpha} = 0.9 \, \text{a.u.}^{-1}$, $\overline{\theta_0} = -5$, $\overline{\zeta} = 2$, $\overline{\kappa} = 0.5$, $\beta = 0.05$, with $10^3$ simulations per model.

Figs~\ref{fig:param_mo} to~\ref{fig:env_mo} show the collective behavior metrics for different parameters and experimental conditions; the grand mean accuracy ($Q$), the grand mean residence time in the best patch ($T_\text{res}$), and the mean absolute derivative over time of $\overline{q}$ ($D$). The variant $D_\mathrm{group}$ shown in \ref{fig:D_int}~Fig and \ref{fig:D_int_env}~Fig is the mean over time and over groups of the absolute derivative of group accuracy $q$.

They were calculated from a mean over the parameter space (except for the one parameter per subplot which varies in Fig~\ref{fig:param_mo}), with 50 simulations per parameter set. Error bars are omitted for clarity.

\subsubsection{Model selection and parameter identifiability}

We used two identification methods: \textbf{(1) Bayesian inference} with Monte Carlo likelihood approximation and maximum a posteriori estimation~\cite{youngEssentialsStatisticalInference2010,gelmanBayesianDataAnalysis2013}, and \textbf{(2) Wasserstein distance minimization}~\cite{kantorovichTranslocationMasses1942,Vaserstein1969,villaniTopicsOptimalTransportation2003}. We selected models (NI, CT, CB, PT, PB) and assessed parameter identifiability using two summary statistics: the distribution of group accuracy (fraction of agents in the highest-quality patch) at multiple sampling times, or the distribution of departure times. For each experimental condition, model, and parameter set (see above for the parameter space), we generated 600 simulations to obtain distributions of group accuracy and departure times, each bin corresponding to a sampling time. These numerical distributions were then compared to synthetic data. When not stated otherwise,
the number of simulations for these in silico observations is 20. We quantified success as the proportion of correct model ($\mathcal{M}$) and mean parameter ($\mathcal{P}$) recovery across 400 repeated inference runs.

Please note that the models are of similar complexity and share the same number of parameters, except for the NI model, which lacks the parameter $\overline{\kappa}$. Because this project focuses on the effect of experimental conditions on identification, and because maximum a posteriori estimation yielded better identification results than the other methods we tested, we relied on this approach here. Future work should compare models using information criteria such as the AIC (see the Discussion).

\subsubsection{Default experimental conditions} Unless otherwise specified, the simulations were carried out with the reward probability in the best patch $p^1 = 0.8$, the reward probability ratio $p^0/p^1 = 0.2$, the group size $N = 4$, the initial group accuracy $Q_\mathrm{in} = 0$, the travel time $\tau_\text{tr} = 0.2 \, \text{a.u.}$, the trial duration $50 \, \text{a.u.}$, and the sampling period $1$ a.u.

\section{Results}
\subsection{Characterization of the socio-cognitive models}

We began by characterizing how the five models; NI, CT, CB, PT, and PB, generate distinct dynamical signatures across parameters and conditions. This foundational analysis enabled the interpretation of model selection and parameter identifiability in the subsequent sections.

\subsubsection{Collective dynamics}
\begin{figure}[!h]
    \centering
\includegraphics[width=0.95\linewidth]{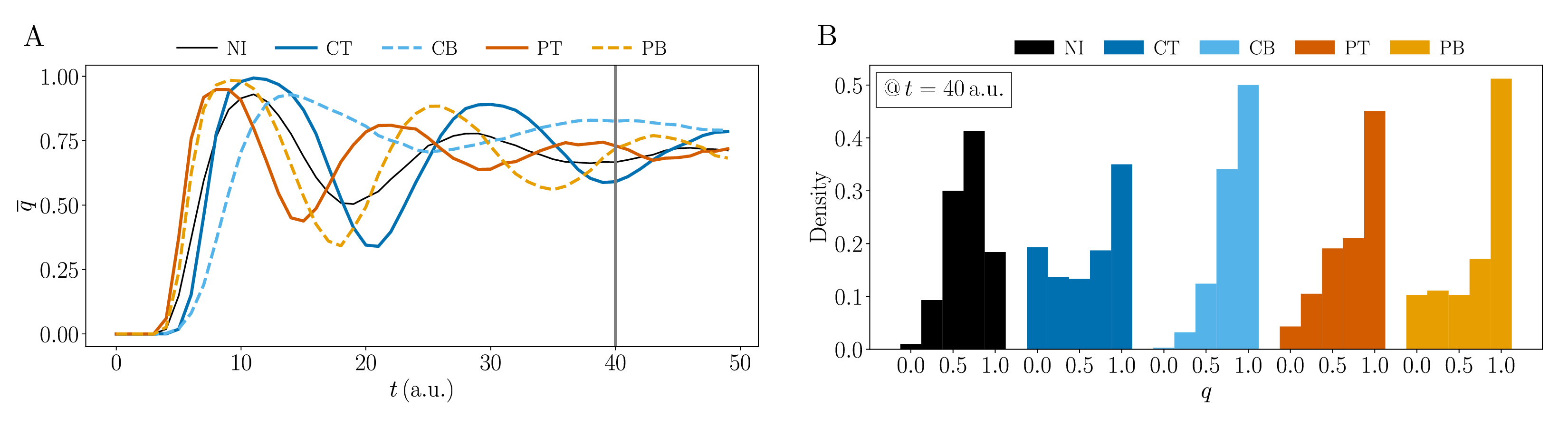}
    \caption{\textbf{Group accuracy of the five socio-cognitive processes.}: NI (black, solid line), CT (dark blue, solid line), CB (light blue, dotted line), PT (dark orange, solid line), PB (light orange, dotted line). (A) Mean accuracy $\overline{q}$ as a function of time. (B) Distributions of group accuracy $q$ at $t = 40$ a.u.}    
    \label{fig:dyn1}
\end{figure}

All models exhibit a damped oscillatory mean accuracy $\overline{q}$ (Fig~\ref{fig:dyn1}A). The agents initially leave patch $k=0$ for patch $k=1$, then return cyclically. The dampening reflects progressive desynchronization as individuals are more distributed across the two patches, eventually approaching a steady-state distribution.

Please note that the damping can have two different causes: each group's oscillation decays, i.e. individuals during the same trial are more distributed, or dephasing, i.e. each group keeps oscillating, but their phases drift apart, so the average looks damped. We see that the mean rate of change of group accuracy as a function of time in \ref{fig:damp}~Fig follows the pattern of the oscillations observed in Fig~\ref{fig:dyn1}, highlighting that these oscillations are not just the mere product of group dephasing.

Social information clearly alters these dynamics by shifting oscillation peaks, changing the accuracy, and modifying the degree of dampening. These three features reveal different aspects of group behavior: exploitation, group success, and synchronization. We quantify these features using three complementary metrics: grand mean residence time $T_\text{res}$ (characteristic timescale of patch switching), and two metrics directly from the mean accuracy $\overline{q}$; grand mean accuracy $Q$ (time-averaged fraction of agents in the best patch), and mean magnitude of the time derivative of $\overline{q}$, $D$. These metrics will serve as the basis for further model characterization in Figs~\ref{fig:param_mo} to~\ref{fig:env_mo}. For clarity reasons, in the rest of the study, we will use the shorter terms \emph{residence time} for $T_\text{res}$, \emph{accuracy} for $Q$, and \emph{rate of change} for $D$.

Counting and pulsatile representations produce opposite temporal shifts: compared to the non-interacting model, the counting (resp. pulsatile) representations lead to right (resp. left) shifted oscillations, corresponding to delayed (resp. earlier) departures.

Compared to the non-interacting model, the CT, PT, and PB models exhibit increased oscillation amplitude and similar accuracy value, consistent with enhanced rate of change. In contrast, CB produces damped oscillations and elevated accuracy, suggesting that continuous belief updating suppresses coordinated exploration movements in favor of sustained exploitation.

Fig~\ref{fig:dyn1}B presents the distributions of group accuracy $q$ at fixed time $t=40$ a.u. These distributions, sampled at multiple time points, serve as the summary statistics for model selection in Section~\ref{sec:id} below.

\subsubsection{Quantification of the dynamics along the parameter space}

\begin{figure}[!h]
    \centering
    \includegraphics[width=0.95\linewidth]{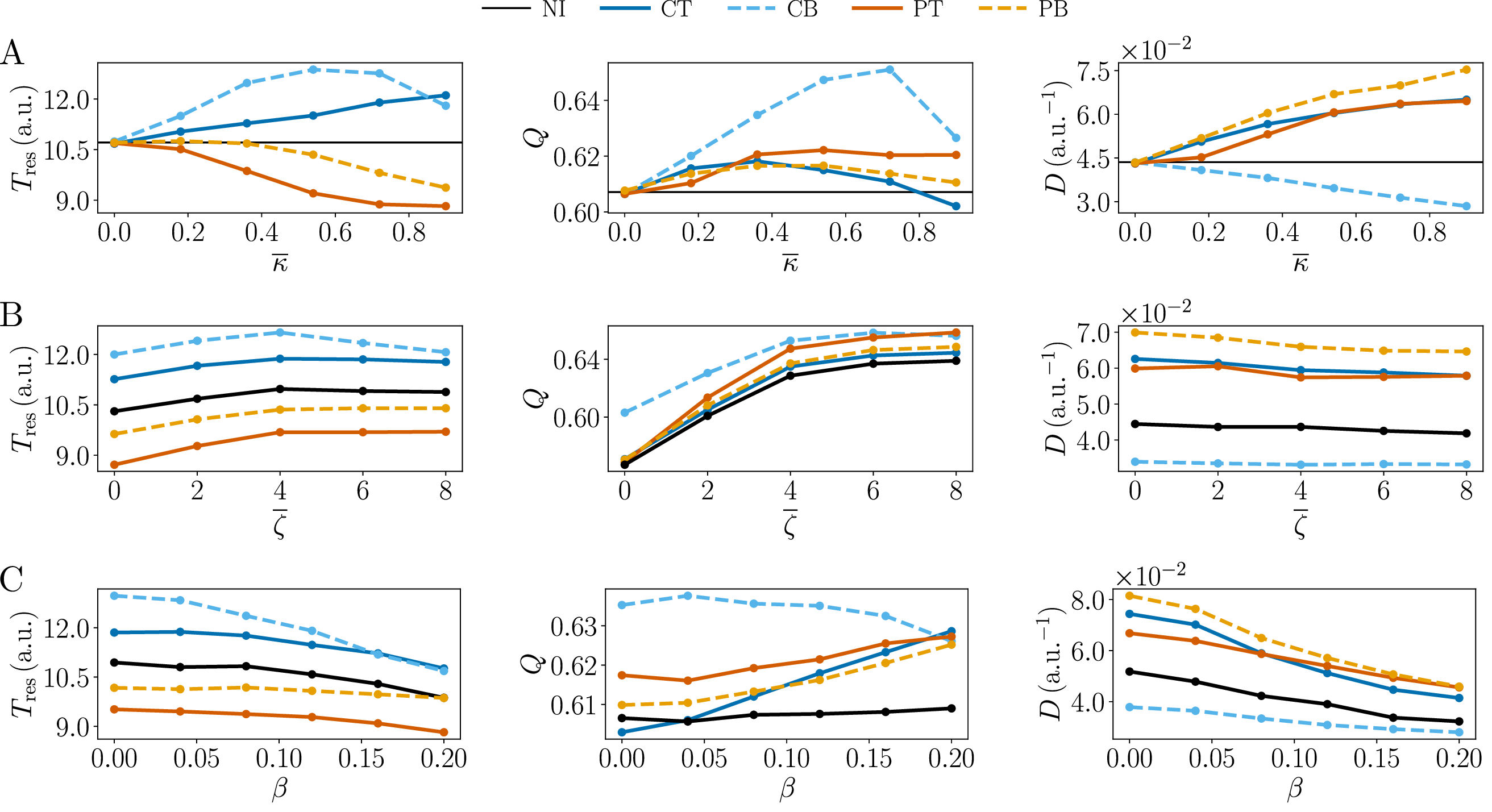}   
    \caption{\textbf{Effects of model parameters on group behavior metrics.} Residence time (left), accuracy (middle), and rate of change (right) for the five socio-cognitive models: NI (black solid line), CT (dark blue solid line), CB (light blue dotted line), PT (dark orange solid line), PB (light orange dotted line), plotted as a function of (A) social coupling strength $\kappa$, (B) learning bias strength $\zeta$, and (C) individual heterogeneity $\beta$.}    
    \label{fig:param_mo}
\end{figure}

The classical parameters of the drift-diffusion model ($\overline{\alpha}$, $\overline{\theta_0}$ and $\overline{\sigma}$) have well-established effects on decision-making dynamics~\cite{ratcliffDiffusionDecisionModel2016}. Therefore, we focus on the three other individual parameters: the social coupling strength $\overline{\kappa}$, the learning bias strength $\overline{\zeta}$, and the degree of individual heterogeneity $\beta$. Fig~\ref{fig:param_mo} displays how these parameters modulate the three key metrics ($T_\text{res}$, $Q$, $D$) across all five models.

\paragraph{Social coupling strength}
Figure~\ref{fig:param_mo}A reveals contrasting effects of the representation of social information on the oscillation timing. Counting representations (CT, CB) lengthen residence time $T_\text{res}$, consistent with agents interpreting high neighbor density as a signal of patch quality and sustained exploitation. Conversely, pulsatile representations (PT, PB) shorten $T_\text{res}$, suggesting that discrete arrival/departure signals promote rapid exploration.

Accuracy $Q$ shows different behaviors across modulation modes: CB alone increases $Q$ substantially, reflecting sustained allocation to the best patch, while CT, PT, and PB remain near NI levels. This elevation in CB is robust, not merely a phase shift of oscillations; in contrast, the smaller fluctuations in CT, PT, and PB around the NI baseline reflect primarily temporal shifts in oscillation peaks rather than changes in steady-state exploitation.

The rate of change ($D$) increases sharply for pulsatile models (PT, PB), indicating tightly coordinated, high-amplitude switching events; agents collectively abandon patches in response to discrete departure signals. CB shows the opposite trend, with reduced $D$ and prolonged residence times, indicating a damped oscillatory regime where agents remain longer in high-quality patches. This decoupling of metrics is instructive: high $T_\text{res}$ paired with low $D$ (CB) signals sustained exploitation with gradual transitions, whereas low $T_\text{res}$ (PT, PB) with high $D$ signals rapid, synchronized cycling. The CT model is more ambiguous, as it shows long residence times \emph{and} important oscillations. \ref{fig:D_int}A~Fig shows that the intra-group rate of change of CT is lower than for models with pulsatile representations (but still higher than CB), suggesting that the oscillations of $\overline{q}$ are also the result of a dephasing between the groups. The CT model thus yields a trade-off dynamics, between exploration and exploitation.

\paragraph{Learning}
Figure~\ref{fig:param_mo}B shows that learning information (encoded in initial bias $z_i^k$) consistently enhances collective performance. Increasing $\overline{\zeta}$ lengthens $T_\text{res}$, showing that agents with stronger memory of past patch quality sustain exploitation longer. Accuracy $Q$ increases across the board, demonstrating that learning amplifies the ability to track and exploit the best patch. Rate of change $D$ remains relatively stable, suggesting that learning primarily strengthens exploitation efficiency without altering the group synchronization.

\paragraph{Inter-individual variability}
Figure~\ref{fig:param_mo}C demonstrates that parameter heterogeneity, captured by the coefficient of variation $\beta$, shortens $T_\text{res}$ across all models, likely because agents with low drift ($\alpha_i$), high threshold ($\theta_{0,i}$) or weak bias ($\zeta_i$) abandon patches more readily, setting the timescale for the group. Accuracy $Q$ exhibits a divergent response: All models but CB improve with heterogeneity, suggesting that diversity enables exploration of alternative behavioral strategies that enhance collective success. CB, by contrast, deteriorates with increasing $\beta$, indicating that heterogeneity disrupts the sustained exploitation regime that defines this model's strength. Rate of change $D$ decreases with increased heterogeneity, while intra-group rate of change is stable (and can also slightly increase), see \ref{fig:D_int}C~Fig. Parameter variability then drives groups apart, but during the same trial, individuals are similarly cohesive.

\subsubsection{Quantification of the dynamics for varying experimental conditions}
\begin{figure}[!h]
    \centering
    \includegraphics[width=0.95\linewidth]{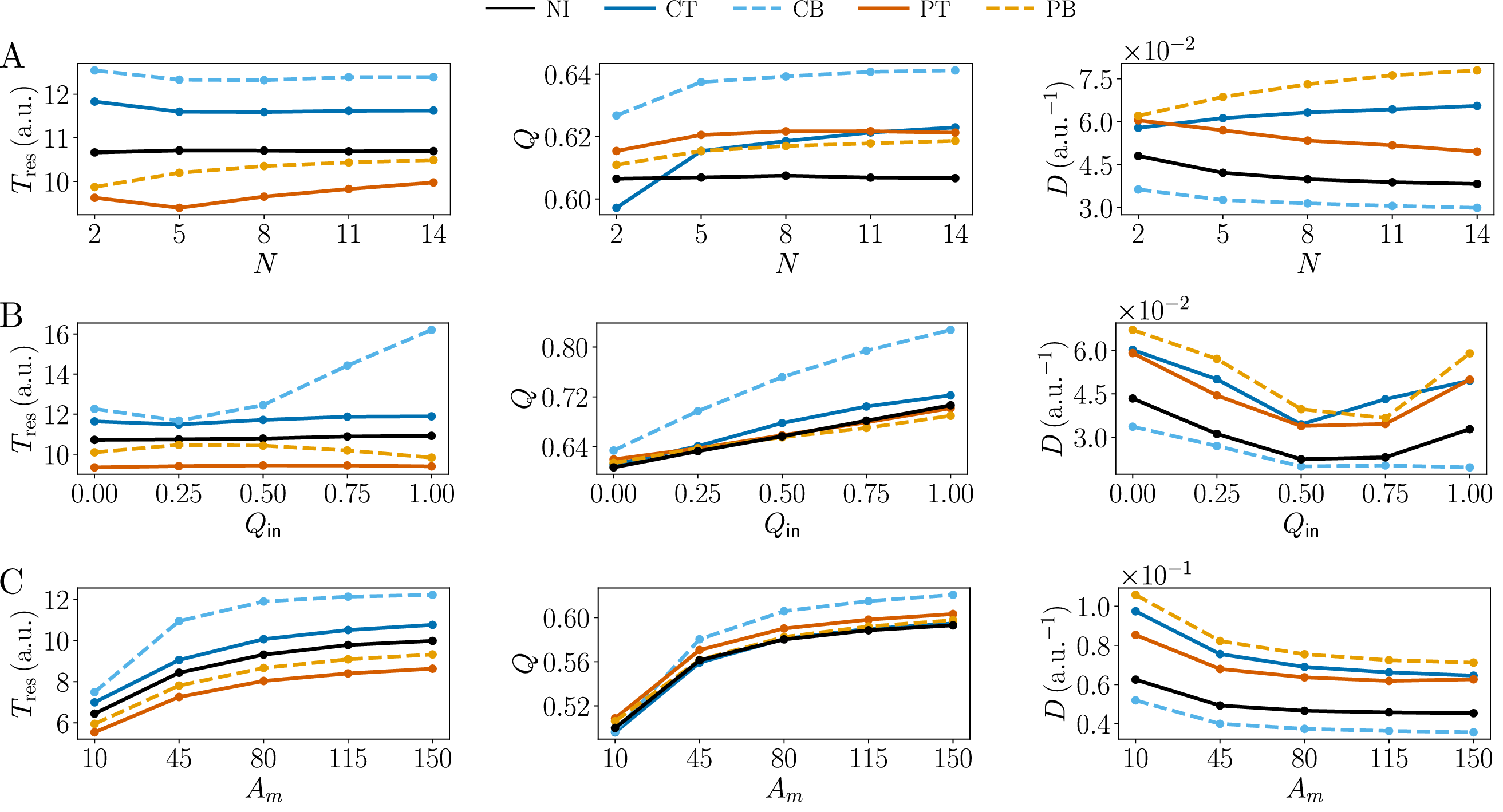}
    \caption{\textbf{Effects of experimental conditions on group behavior metrics.} Residence time (Left), accuracy (Middle) and rate of change (Right), of the five socio-cognitive processes: NI (black solid line), CT (dark blue solid line), CB (light blue dotted line), PT (dark orange solid line), PB (light orange dotted line), plotted as a function of (A) the group size $N$, (B) the initial group accuracy $Q_\mathrm{in}$, and (C) the resource depletion parameter $A_m$.}    
    \label{fig:exp_mo}
\end{figure}

To further characterize the dynamics of these cognitive models, we studied how the experimental conditions modulate the five socio-cognitive models by systematically varying three experimental parameters; group size ($N$), initial group distribution ($Q_\text{in}$), and resource depletion ($A_m$). Fig~\ref{fig:exp_mo} displays the resulting trajectories of residence time $T_\text{res}$, accuracy $Q$, and rate of change $D$.

\paragraph{Group size}
Residence time increases for pulsatile models (PT, PB) as group size grows, see Fig~\ref{fig:exp_mo}A, reflecting stronger collective signals from more departure events. Counting models (CT, CB) show initial decreases followed by plateaus, suggesting that density signals saturate at moderate group sizes. Accuracy improves across all social models, especially for counting models, with increasing $N$, remaining flat for NI, indicating that social information scales cooperatively with group size. Rate of change $D$ decreases with $N$ for NI, PT, CB and increases for CT and PB, revealing non-trivial interactions between the social information representation and modulation mechanisms. \ref{fig:D_int}D~Fig shows that for all models, inside groups individuals become less cohesive with $N$ (decreasing $D_\mathrm{group}$). Indeed, one decision from one individual becomes then a less important observation compared to the whole group.
Then, the increase of $D$ with $N$ for CT and PB corresponds to increased inter-group variability.

\paragraph{Initial agents distribution}
Fig~\ref{fig:exp_mo}B shows that across most models, residence time is stable under different initial fractions of agents in the best patch $Q_{\text{in}}$, except for CB which shows substantially longer residence times at high $Q_{\text{in}}$: because agents remain in the best patch once established, initialization in that patch amplifies this stickiness and extends overall residence. 
Accuracy increases monotonically with $Q_\text{in}$, reflecting the short trial duration where initial conditions dominate the dynamics. The evolution of $D$ displays an U-shape tendency, with polarized initial conditions (e.g., $Q_\text{in} = 0$ or $Q_\text{in} = 1$) inducing larger oscillations because the group begins far from its steady-state, so it takes more time to reach it.

\begin{figure}[!h]
    \centering
    \includegraphics[width=0.7\linewidth]{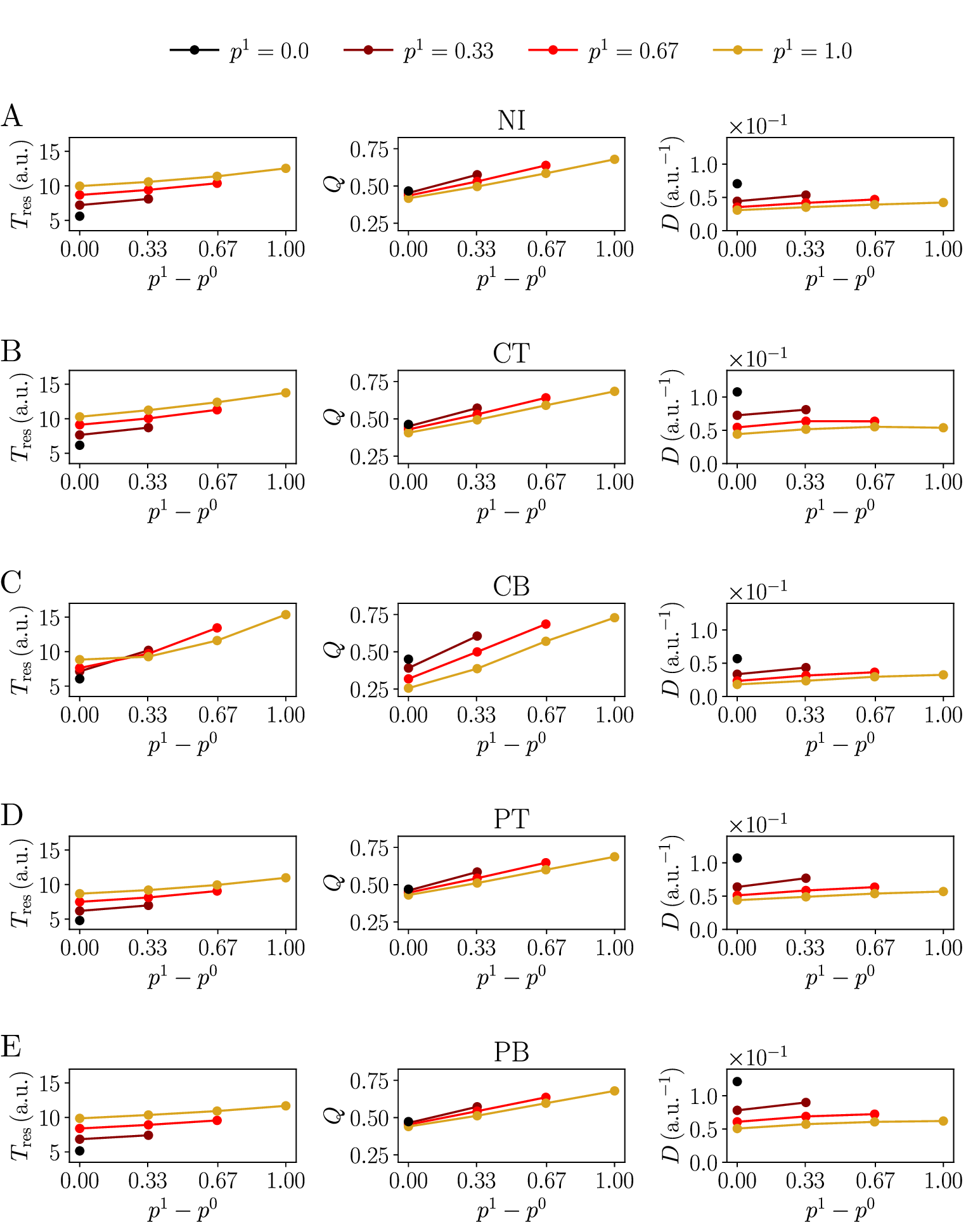}
    \caption{\textbf{Effects of the resource distribution on group behavior metrics.} Residence time (Left), accuracy (Middle) and rate of change (Right), of the five socio-cognitive processes (from left to right): (A) NI, (B) CT, (C) CB, (D) PT, and (E) PB, as a function of the difference between reward probabilities $p^1-p^0$, for different $p^1$ values, see color legend on top.}
    
    \label{fig:env_mo}
\end{figure}

\paragraph{Resource depletion}
Lower maximum resource capacity $A_m$ values correspond to stronger depletion: patches exhaust faster, and reward probabilities decline more steeply. Across all models, Fig~\ref{fig:exp_mo}C shows that depletion reduces residence time, accuracy, and increases rate of change. The effects are mechanistically transparent: smaller $A_m$ accelerates the transition toward low, indistinguishable reward probabilities. Low reward probabilities lead to shortened residence times (agents encounter fewer rewards and thus accumulate weaker evidence for patch quality), and increased rate of change, see Fig~\ref{fig:env_mo}. Accuracy declines because patches become increasingly similar in their reward statistics, reducing the information content available to discriminate between them.

\paragraph{Resource distribution}
Fig~\ref{fig:env_mo} shows that increasing the difference between reward probabilities of each patch, $p^1 - p^0$, increases residence time and accuracy across all models: dissimilar patches favor exploitation of the best patch. For a given $p^1-p^0$, larger $p^1$ increases residence time, but decreases accuracy. Rate of change depends primarily on the $p^1$ value, increasing with smaller $p^1$, associated with more coordinated departures of less rewarding patches. Interestingly, CB exhibits heightened sensitivity to resource distribution in residence time and accuracy than other models, but less sensitivity for the rate of change.\\

Having characterized the five models' collective dynamics and how they are modulated by different experimental conditions, we then asked whether these distinct signatures are sufficiently stable and separable to enable reliable model selection and parameter identifiability from behavioral data.

\subsection{Model selection and parameter identifiability}
\label{sec:id}
\subsubsection{Identification across metrics and estimation methods}

\begin{figure}[!h]
    \centering
    \includegraphics[width=0.7\linewidth]{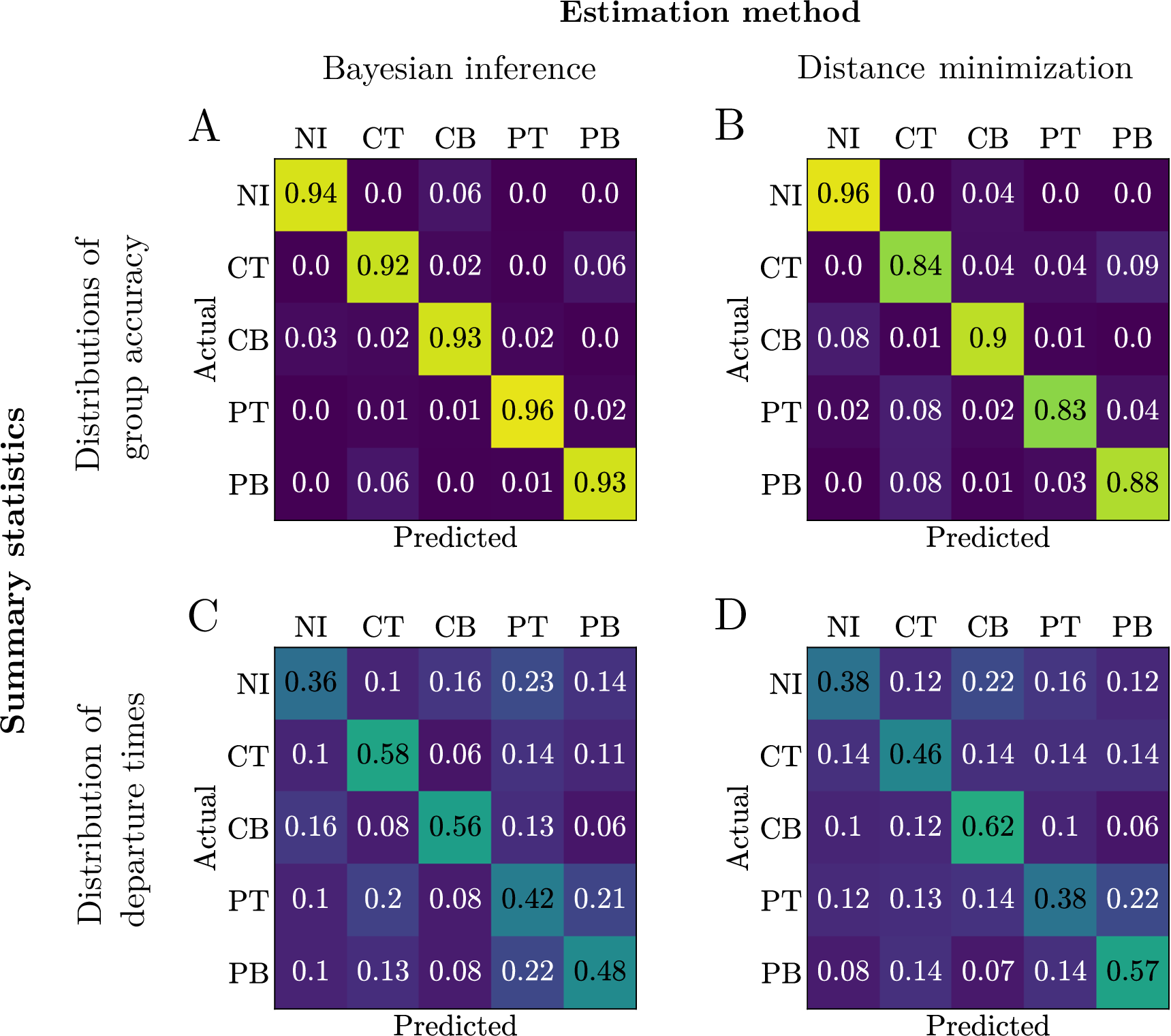}
    \caption{\textbf{Confusion matrices for the five socio-cognitive models} (NI, CT, CB, PT, PB), using distributions of accuracy at different sampling times as a summary statistic (Top; A, B) or using the distribution of departure times (Bottom; C, D), obtained with Bayesian inference (Left; A, C) and distance minimization (Right; B, D). Each entry shows the fraction of trials predicting model $\mathcal{M}$ given data generated from the labeled model.}
    \label{fig:confM}
\end{figure}

Both Bayesian inference and distance minimization achieve substantially higher identification accuracy using distributions of accuracy as the summary statistic compared to distribution of departure times, see Fig~\ref{fig:confM}. The distribution of departure times exhibits a discrete, noisy structure that obscures model differences at low observation counts; reliable distinction between models requires substantially more data. Given this fundamental limitation, we focus on accuracy as the primary metric for subsequent identification analyses.\\

\begin{figure}[!h]
    \centering  
    \includegraphics[width=0.75\linewidth]{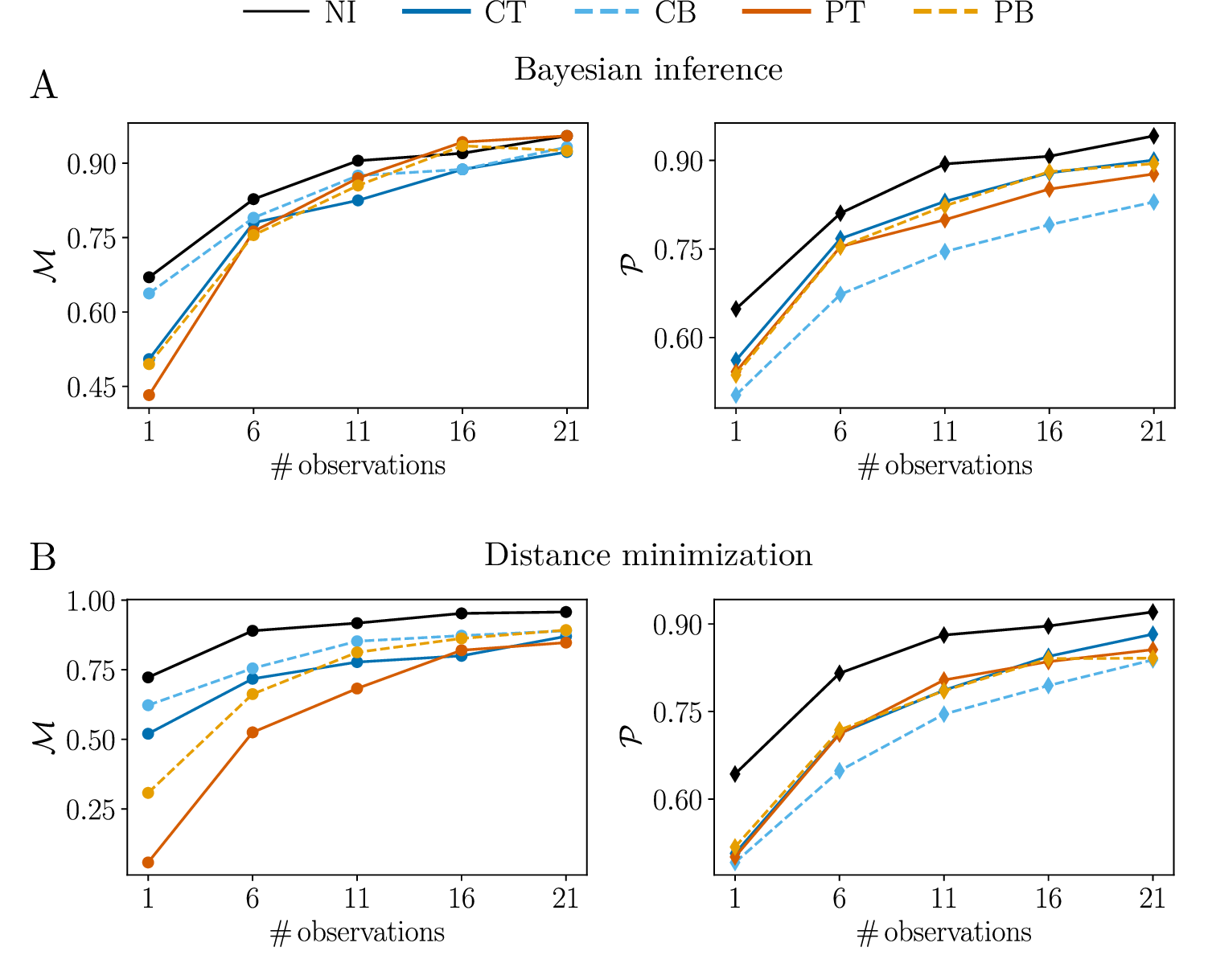}
    \caption{\textbf{Model selection and parameter identifiability success as a function of observation count}, using Bayesian inference (Top) and distance minimization (Bottom). Each panel shows the success rate for the five socio-cognitive models: fraction of trials correctly identifying models $\mathcal{M}$ (Left) or parameters $\mathcal{P}$ (Right).}
    \label{fig:obs_id}
\end{figure}

Fig~\ref{fig:obs_id} demonstrates that prediction accuracy for both model and parameter estimation improves with sample size for both methods, with Bayesian inference substantially outperforming at small sample sizes. We then focus here on the results for Bayesian inference. For small samples, pulsatile representations exhibit poorer model selection compared to counting representations. This discrepancy suggests that pulsatile models are more sensitive to limited observations. We hypothesize that pulsatile models, as they are associated with short residence times, produce noisy signatures in small samples (lot of departures), but then generate more informative features as sample size increases. The parameter estimation success follows the ranking CB $<$ PT $<$ PB $\approx$ CT $<$ NI. Please note that NI achieves the highest performance for both model and parameter prediction. We attribute this superior performance to two factors: (1) the absence of social coupling reduces sensitivity to variability, and (2) fewer parameters need to be estimated (no $\kappa$).  Except NI, the parameter identifiability observations correlate with accuracy and rate of change metrics (see Fig~\ref{fig:param_mo}A), suggesting that oscillatory dynamics and different accuracy values provide richer information for parameter extraction. Further analysis in various conditions will provide more precise insights on the causes of identification efficiency.

\subsubsection{Identification across experimental conditions}

\begin{figure}[!h]
    \centering  
    \includegraphics[width=0.75\linewidth]{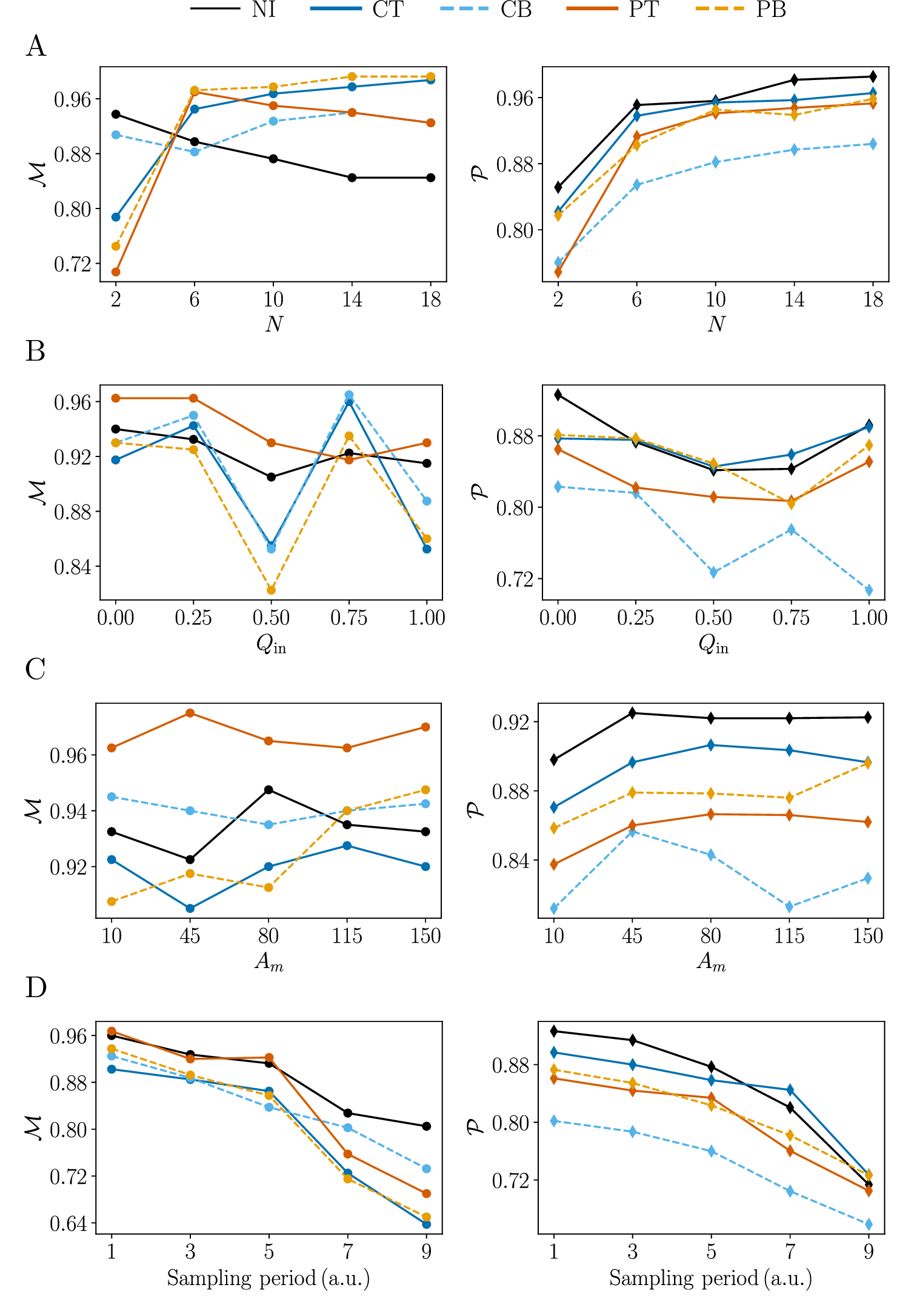}
    \caption{\textbf{Effects of experimental conditions on model selection and parameter identifiability.} The proportion of model $\mathcal{M}$ (Left) and parameter $\mathcal{P}$ (Right) correctly predicted between the five socio-cognitive models; NI (black solid line), CT (dark blue solid line), CB (light blue dotted line), PT (dark orange solid line), PB (light orange dotted line), are obtained with Bayesian inference. They are plotted as a function of (A) the group size $N$, (B) the initial group accuracy $Q_\mathrm{in}$, (C) the depletion parameter $A_m$, and (D) the sampling period.}
    \label{fig:exp_id}
\end{figure}

\paragraph{Group size}
Model selection in Fig~\ref{fig:exp_id}A shows distinct behavior across two regimes. For a small group size ($N = 2$), limited observations lead to poor model selection for PT, PB, and CT, while CB and NI show superior performance. This pattern mirrors the small-sample regime observed in Fig~\ref{fig:obs_id}. For larger group sizes ($N \geq 6$), model selection exhibits differential sensitivity to group size. For CT and PB, it improves with increasing $N$, while it declines with PT and NI. Interestingly, this mirrors the evolution of the rate of change $D$ with $N$ in Fig~\ref{fig:exp_mo}A. CB exhibits a non-monotonic relationship with $N$ ($\mathcal{M}$ increasing then decreasing), suggesting that model selection for CB results from complex interactions between its features, particularly given that it exhibits less oscillatory behavior than other models.

Parameter identifiability improves with $N$ across all models in a consistent manner, with the most substantial gains occurring between $N = 2$ and $N = 6$. We hypothesize that this improvement reflects an increase in total observations, as larger groups provide more data points overall. We note that here also the ranking of parameter identifiability success ($\mathrm{CB} < \mathrm{PT} < \mathrm{PB} < \mathrm{CT} < \mathrm{NI}$) correlates with the ordering of accuracy metrics (see Fig~\ref{fig:exp_mo}A).

\paragraph{Initial agents distribution}
Model selection success $\mathcal{M}$ exhibits an M-shaped dependence on the initial fraction of agents in the best patch $Q_{\text{in}}$ (Fig~\ref{fig:exp_id}B). Identification is poorest at fifty-fifty initial conditions ($Q_{\text{in}}=0.5$), improves toward extreme configurations ($Q_{\text{in}}=0$ or $Q_{\text{in}}=1$), and reaches optimal performance at intermediate values ($Q_{\text{in}}=0.25$ or $Q_{\text{in}}=0.75$). This non-monotonic pattern reflects the interplay of two mechanisms. First, we see a parallel with the rate of change $D$ (see Fig~\ref{fig:exp_mo}B), which exhibits reduced oscillations away from polarized initial conditions, suggesting that oscillatory dynamics enhance model selection. However, this alone does not account for the superior performance at intermediate values. At these conditions, oscillations remain pronounced while agents are distributed across patches in diverse configurations, thereby providing richer and more discriminative dynamical signatures that facilitate model discrimination.

Parameter identifiability success $\mathcal{P}$ follows a U-shaped dependence on $Q_{\text{in}}$, paralleling the pattern observed in rate of change $D$. This correspondence underscores the importance of oscillatory dynamics for accurate parameter extraction: oscillations provide the temporal variability and feature richness necessary to distinguish parameter values.

\paragraph{Resource depletion}
Model selection shows minimal sensitivity to resource depletion in Fig~\ref{fig:exp_id}C (increased depletion for smaller $A_m$ parameter, the maximum amount of food per patch), with performance fluctuating around a relatively constant baseline across depletion levels.

In contrast, parameter identifiability success $\mathcal{P}$ exhibits a modest decline as depletion increases. This pattern closely parallels the dependence of accuracy $Q$ on $A_m$ (Fig~\ref{fig:exp_mo}C), suggesting that group-level accuracy is a primary determinant of parameter recovery. 

\paragraph{Sampling period}
Increasing the sampling period decreases both model selection and parameter identifiability success across all models (Fig~\ref{fig:exp_id}D), as expected. The degradation in model selection $\mathcal{M}$ is more pronounced for the oscillating models (PT, PB, CT) than for CB, reinforcing the critical role of oscillatory dynamics in model discrimination. This differential sensitivity reflects the distinct mechanistic signatures of these models: PT, PB, and CT are primarily characterized by their oscillatory behavior, whose temporal structure is readily obscured by coarse temporal sampling. Conversely, CB relies less on oscillatory signatures and is instead distinguished by its elevated accuracy $Q$, a feature that degrades more gracefully under temporal undersampling. This observation underscores a fundamental principle: model selection depends on preserving the dynamical features most diagnostic of each model. For oscillation-driven models, fine temporal resolution is essential; for accuracy-driven models, coarser sampling poses less identification risk. 

\begin{figure}
    \centering
    \includegraphics[width=0.7\linewidth]{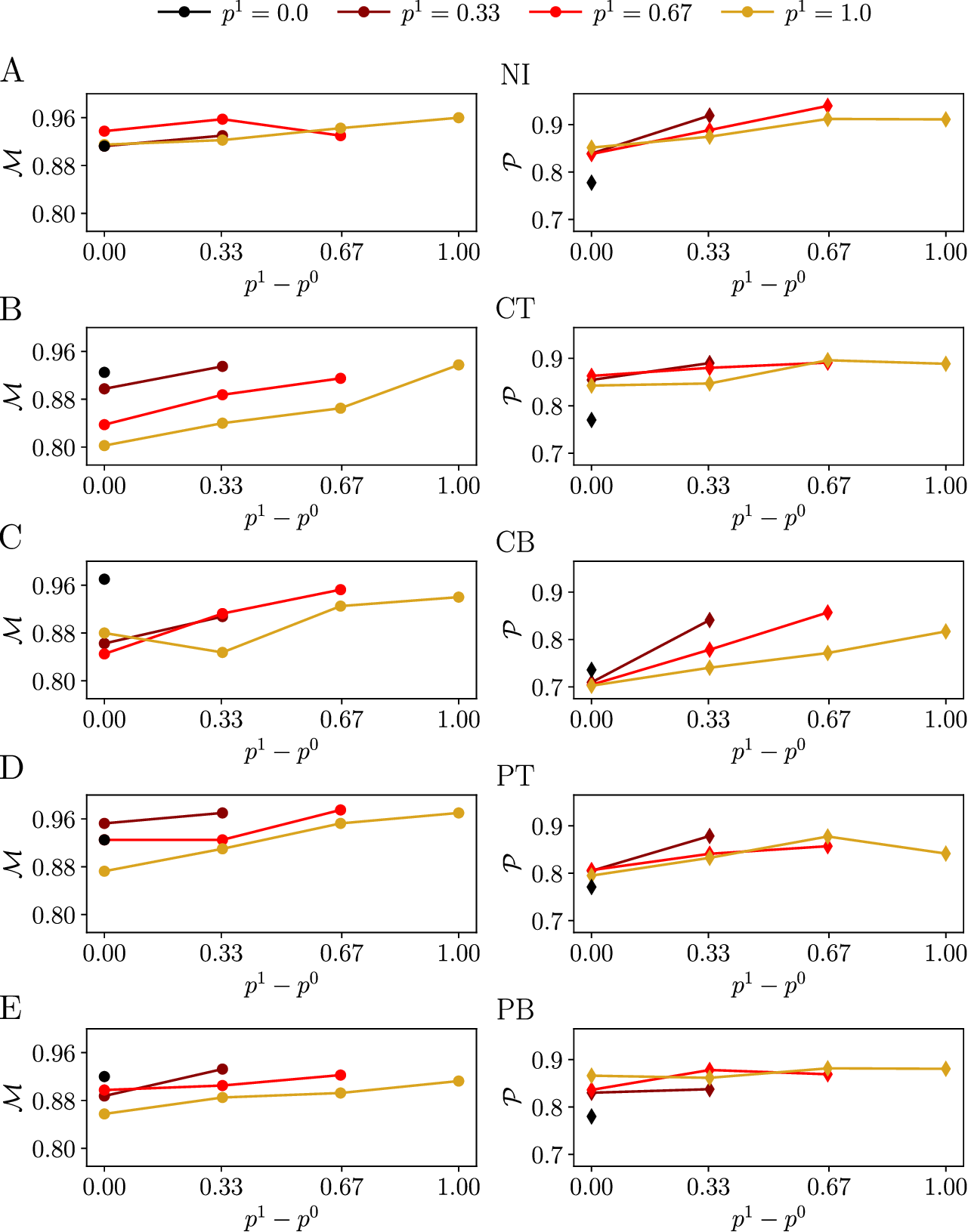} 
    \caption{\textbf{Effects of resource distribution on model selection and parameter identifiability.} The proportion of correctly identified model ($\mathcal{M}$, left column) and correctly recovered parameters ($\mathcal{P}$, right column), are shown across the five computational models: (A) NI, (B) CT, (C) CB, (D) PT, (E) PB, obtained via Bayesian inference, as a function of the difference between reward probabilities $p^1-p^0$, for different $p^1$ values, see color legend on top.}
    \label{fig:env_id}
\end{figure}

\paragraph{Resource distribution}
Fig~\ref{fig:env_id} shows how model selection and parameter identifiability succeed across different resource distributions. This figure is complex to read, but we can nonetheless extract some general trends. Both $\mathcal{M}$ and $\mathcal{P}$ increase with larger $p^1 - p^0$ and decrease with $p^1$, and both correlate with behavioral metrics, accuracy $Q$ and rate of change $D$ (see Fig~\ref{fig:env_mo}). This correspondence is particularly pronounced for two models: $\mathcal{M}$ for CT scales closely with $D$, and $\mathcal{P}$ for CB scales closely with $Q$. This dissociation reinforces the idea that parameter recovery relies somewhat more on accuracy ($Q$), while model selection relies somewhat more on rate of change ($D$).

Apart from these main tendencies, not all models follow this pattern cleanly, pointing to more complex underlying interactions. In particular, $\mathcal{P}$ for PB does not track $Q$ as the other models do, but instead scales like $T_\text{res}$, with high-$p^1$ lines lying above rather than below, the opposite sign to the main $\mathcal{P}$-$Q$/$D$ correspondence, suggesting that model-specific mechanisms not captured by the general trend can also shape identification.

Special attention should be paid to the degenerate case $p^0 = p^1 = 0$, where $\mathcal{P}$ is markedly degraded. This is readily interpretable: with no contrast and no reward at all, the learning-related parameter $\overline{\zeta}$ has no effect on behavior, its associated term vanishes, and so cannot be correctly recovered from the data (see \ref{tab:zeta_worse}~Table).

These results provide general guidelines for environment design: efficient model selection and parameter identifiability are favored by dissimilar patches (large $p^1 - p^0$) and, for a given gap, lower absolute reward probabilities, while the degenerate case $p^0 = p^1 = 0$ should be avoided, as it specifically compromises parameter recovery.

\section{Discussion}
How social information is represented and integrated into individual decisions and what collective behavior this produces is a central question for understanding the emergence and causal structure of collective systems~\cite{couzinCollectiveCognitionAnimal2009}. Addressing this question moves us beyond observations that groups behave differently than isolated individuals, to move towards generative hypotheses.

Indeed, the underlying cognitive computations are typically hidden from direct observation. This is why combining experiments with computational models is so valuable: experiments generate data, and models formalize candidate mechanisms, translating observed patterns into explicit rules for how individual decision-making gives rise to collective outcomes. Models generate predictions that experiments can then validate or invalidate; discrepancies between model predictions and experimental results point either to flawed mechanisms in the model or to boundary conditions the experiment did not capture, and iterating between the two improves understanding.

Building on this logic, we asked whether such mechanisms can be recovered from the macroscopic dynamics they produce. Focusing on location cues in patch-foraging tasks, we distinguished two dimensions along which social information processing may vary: how information is represented; continuous estimation of the social density (counting representation) versus discrete monitoring of conspecific arrivals or departures (pulsatile representation), and how it modulates decision behavior; by shifting the decision threshold (threshold modulation) or by contributing directly to evidence accumulation (belief modulation).

This question is made difficult by the state of existing model selection and parameter identifiability methods. While identification analysis is well established for deterministic, ODE-based models, it remains considerably more difficult for stochastic, nonlinear models, for which identification methods typically must be adapted anew for each modeling framework~\cite{chisStructuralIdentifiabilitySystems2011,villaverdeStructuralIdentifiabilityDynamic2016,hasanIdentifyingLatentStochastic2022,aguilarLimitsInferenceComplex2026}. Our study contributes to this literature by providing a systematic identification analysis for a class of socio-cognitive foraging models, and by relating identification directly to interpretable features of collective behavior.

\subsection{Collective behavioral signatures of the cognitive models display parallels with model selection and parameter identifiability}
We first consider what made identification more or less efficient at a methodological level, before turning to the broader mechanistic picture that emerged from our results.

\paragraph{Identification methods and summary metrics} Model selection and parameter identifiability performed better when using group accuracy as a summary statistic rather than the distribution of departure times, likely because the former is a smoother, less discrete measurement~\cite{gelmanBayesianDataAnalysis2013}. Bayesian inference also outperformed distance minimization for small samples (Fig~\ref{fig:confM}). 

\paragraph{Parallels between behavioral metrics and identification} Beyond the choice of summary statistic and method, a central theme emerged from our analysis: \emph{model selection and parameter identifiability are not intrinsic, fixed properties of a given socio-cognitive mechanism, but instead closely track the structure of its behavioral signature}. Throughout our characterization of collective foraging dynamics, three metrics; residence time ($T_\text{res}$), accuracy ($Q$), and rate of change ($D$), consistently paralleled the identification measurements $\mathcal{M}$ (model selection efficiency) and $\mathcal{P}$ (parameter identifiability efficiency) across parameter regimes and experimental manipulations. This convergence provides a mechanistic account of why identification succeeds or fails under a given condition, rather than a purely descriptive one, and suggests that these behavioral metrics constitute an interpretable bridge between collective dynamics and identification.

Across experimental conditions, $D$ emerged as the behavioral correlate most consistently associated with $\mathcal{M}$: for example as a function of $N$ (Fig~\ref{fig:exp_mo}A and Fig~\ref{fig:exp_id}A) or under different resource distributions (Fig~\ref{fig:env_mo} and Fig~\ref{fig:env_id}), parallels can be drawn between the behavioral metric and the identification measurement. Model selection was degraded when oscillatory, synchronized switching was weak.
Furthermore, under small sample sizes (Fig~\ref{fig:obs_id}), small group sizes (Fig~\ref{fig:exp_id}A), and coarser sampling periods (Fig~\ref{fig:exp_id}D), this degradation was systematically more pronounced for the oscillation-driven models (CT, PT, PB) than for CB, whose distinguishing feature is sustained exploitation rather than coordinated cycling. Together, these observations indicate that discriminating how social information is represented and integrated depends largely on the group's capacity to generate temporally resolved, synchronized behavioral transitions, a feature intrinsically sensitive to undersampling.

A complementary pattern held for $\mathcal{P}$ with $Q$: parameter recovery efficiency consistently mirrored the accuracy metric $Q$ across conditions with the ranking of models from the worst to best parameter recovery (Fig~\ref{fig:obs_id}); for example showing a U-shape response to the initial fraction of agents in the best patch (Fig~\ref{fig:exp_mo}B and Fig~\ref{fig:exp_id}B), degrading with resource depletion (Fig~\ref{fig:exp_mo}C and Fig~\ref{fig:exp_id}C), and under different resource distributions (Fig~\ref{fig:env_mo} and Fig~\ref{fig:env_id}). This suggests that the information required for parameter identifiability  is carried primarily by how consistently and effectively the group exploits the better patch, rather than by its synchronization. The break of this correspondence is informative: in Fig~\ref{fig:env_id}, for PB, $\mathcal{P}$ tracked $T_\text{res}$ instead of $Q$, pointing to a model-specific reliance on residence-time information that the general $Q$-$\mathcal{P}$ relationship does not capture, and underscoring that no single behavioral metric is universally sufficient across mechanisms.

The degenerate case $p^0 = p^1 = 0$ offers a clean illustration of this framework's mechanistic value. Here, $\mathcal{M}$ remains comparatively high, consistent with the persistence of informative synchronized dynamics even in the complete absence of reward, while $\mathcal{P}$ collapses, because the learning-related parameter $\overline{\zeta}$ has no behavioral consequence when there is no reward to learn from, and is therefore structurally unrecoverable. 

\paragraph{Broader parallels: simple/complex contagion and artificial systems} These distinctions between representations of social information also resonate with ideas developed outside the foraging literature. The two representations studied here can be related to the distinction between simple and complex contagion in social networks~\cite{centolaCascadeDynamicsComplex2007,centolaHowBehaviorSpreads2018}. Simple contagion is driven by the cumulative number of contacts, akin to the discrete arrival and departure events of the pulsatile representation, whereas complex contagion depends on the proportion of contacts exhibiting a behavior, akin to the density-based signal underlying the counting representation. This parallel would be worth investigating further, as it could help draw connections between group decision-making across different domains, like foraging and opinion dynamics. A related idea also appears in artificial systems, where distinct internal representations can be learned from the same input while supporting comparable behavioral outputs~\cite{liConvergentLearningDifferent2015,lindsayDivergentRepresentationsEthological2021,kumarQuestioningRepresentationalOptimism2025}, echoing our finding that behaviorally similar outcomes can arise from mechanistically distinct representations of social information, and that recovering the underlying representation from behavior alone is a nontrivial identification problem, whether in biological or artificial agents. This distinction also relates to the classical dichotomy between structural and practical identifiability: while the former asks whether parameters are recoverable in principle given the model structure, our results emphasize that practical identifiability is further constrained by the observability and richness of the collective behavioral signature the model generates~\cite{browningIdentifiabilityAnalysisStochastic2020}.\\

Several limitations of our approach should be considered when interpreting these results, spanning the individual-level process, the behavioral metrics we used for interpretation of identification results, and the identification procedure itself.

\subsection{Limitations}
\paragraph{Individual-level process assumptions} At the level of the drift-diffusion process (Eq.~\eqref{eq:xinc}), we assumed white noise, whereas many biological systems exhibit pink noise with long-range temporal correlations. While the white-noise assumption enabled theoretical tractability, it will be important to test whether our predictions hold under more realistic noise processes; we expect such processes to produce enhanced persistence of the drift, wider distributions, and no stable stationary value~\cite{stoyanovPINKNOISE12011}. We also deliberately restricted the strength of the ``concurrent" learning-based threshold modulation, to avoid configurations in which the initial decision variable falls below threshold; stronger learning effects could have been considered.

\paragraph{Behavioral metrics explanatory potential} The metrics we used ($T_\text{res}$, $Q$, $D$) were chosen for their interpretability and their close relationship to the measurements used in the identification analysis, but they are unlikely to be exhaustive. The PB exception in the resource-distribution analysis indicates that additional, model-specific signatures may be needed to fully account for parameter identifiability across all mechanisms. A more precise approach could formally quantify the information content of each metric with respect to $\mathcal{M}$ and $\mathcal{P}$, for example via mutual information or sensitivity analysis. This would allow us to move from the qualitative correspondences described here toward a predictive, quantitative theory of identification in collective behavior models.

\paragraph{Model selection and parameter identifiability technical limits} Our identification analysis relied on a prior calibration to approximate mean departure times (see Methods) and on a coarse grid of parameter values. As expected, \ref{fig:bignl}~Fig shows a decreased identification as a function of $n_l$: this is moderate for model selection, more important for parameter identifiability, but the order of models is globally preserved for the latter,  except for the PT model. This could be explained by the smaller variations of the PT dynamic across $\overline{\kappa}$ values, see Fig~\ref{fig:param_mo}A.

The application of this framework to empirical data requires a verification of these choices, and identification should ideally be re-assessed for the specific experimental conditions used prior to each real laboratory experiment. Our aim here was to provide a general understanding and an accompanying codebase rather than a definitive, experiment-ready pipeline. We also used maximum a posteriori to define the identified model, as it yielded better predictions than the log-likelihood ratio in our setting. It would be important to nonetheless report the log-likelihood ratio as complementary information. Finally, we did not systematically explore how experiment duration affects identification, which would be a valuable addition.\\

Despite these limitations, the parallels we identified between behavioral metrics and identification measurements carry practical consequences that extend beyond the present models, and point toward several directions for future work.

\subsection{Future directions}
\paragraph{Technical improvements} On the technical side, several developments would strengthen the identification protocol underlying our approach. It would be valuable to investigate whether the likelihood of these models can be derived theoretically, at least for the non-interacting case~\cite{liuEfficientInferenceFirst2026}. A theoretical estimate would yield a more robust understanding of the identification process by clarifying the structure of model-parameter relationships. In this study, we compared models of similar complexity (with the exception of the non-interacting model), which simplified model comparison; future work considering models with differing numbers of parameters, for example, separate coupling strengths $\overline{\kappa}$ for arrivals and departures in the pulsatile representation, would require dedicated tools such as Bayes factors, defined as likelihood ratios integrated over parameter values~\cite{gelmanBayesianDataAnalysis2013}, or information criteria such as AIC, BIC, WAIC, or LOO-CV. We note, however, that Bayes factors yielded poorer model selection than maximum a posteriori estimation in preliminary comparisons, a discrepancy that merits further investigation. For more precise parameter estimation, a CMA-ES algorithm~\cite{CovarianceMatrixAdaptationa} could also be combined with distance minimization (we used a grid-based approach in the main text to provide a common framework for comparison with Bayesian inference).

\paragraph{More complex environments} Beyond these methodological considerations, several extensions of the model itself would be valuable. Extending our framework to environments with more than two patches, where a departing agent may move to any of several destinations, would be an important next step, connecting our approach to existing experimental~\cite{robertsForagingRadialMaze1989,franksSpeedCohesionTradeoffs2013} and field~\cite{strandburg-peshkinSharedDecisionmakingDrives2015} work on multi-patch foraging.

\paragraph{Broadening the space of social mechanisms} In a similar spirit, the space of social mechanisms we considered could be broadened. We focused on two representations (pulsatile, counting) and two modulation mechanisms (threshold, evidence accumulation) of social information. Future work could examine additional mechanisms, such as direct active signaling and public information~\cite{dallInformationItsUse2005,leadbeaterSocialLearningInsects2007,galefSocialInfluencesForaging2001}, already studied theoretically in this foraging context~\cite{bidariStochasticDynamicsSocial2022a,blummoyseSocialHierarchyShapes2026}, as well as others: foraging bees, for example, leave chemical marks at exploited patches that other bees learn to avoid~\cite{goulsonForagingBumblebeesAvoid1998}, ants rely on social
information from trophallaxis~\cite{millerSocialInteractionsDiffer2023}, and social cues may sometimes convey only the location of resources without information about their profitability~\cite{hillemannInformationUseForaging2020}. Competitive behaviors~\cite{rantaCompetitionCooperationSuccess1993,lagueEffectsFacilitationCompetition2012} would also be a natural extension, linking our framework to producer-scrounger models~\cite{vickeryProducersScroungersGroup1991} and to non-egalitarian groups in which individuals follow heterogeneous social rules~\cite{scottAnalysisSocialOrganization1956a}. More generally, further work should aim to bring social organization in these models closer to that of real foraging groups~\cite{bakerForagingSuccessJunco1981,leeForagingDynamicsAre2018}. Relatedly, we treated each simulated group as following a single mechanism, whereas real agents may combine multiple mechanisms with individual-specific weights; incorporating such mixtures is an important direction for future models. Finally, since foraging is only one of several behaviors animals engage in during these experiments~\cite{breedAnimalBehavior2016}, and interruptions of foraging can themselves shape collective outcomes~\cite{sumpterCollectiveAnimalBehavior2011,hongResearchResearchingInterrupted2024}, incorporating non-foraging behaviors will be necessary when these models are applied to real data.

\paragraph{Evolutionary comparisons} Beyond these extensions to the model itself, we anticipate that this modeling framework could also serve as a benchmark for evolutionary comparisons across species, opening avenues for quantitative, model-based studies of behavioral evolution~\cite{clarkEvolutionaryAdvantagesGroup1986}.

\paragraph{Application to laboratory experiments} Taken together, these directions point toward a broader goal: this study aims to provide a first bridge between theoretical developments and experimental practice. 

Readily interpretable summary metrics of collective behavior can anticipate where and why identification will be reliable, offering a faster route to experimental design. Our results suggest that observation protocols would benefit of using fine temporal resolution and sufficient group size, and contrastive reward environments.

The methods and accompanying code~\cite{LblummoyseTwopatchforagingidentifiability} could be used by researchers alongside their own experiments, and further extended to assess the model selection and parameter identifiability of other dynamics, socio-cognitive models, and experimental conditions~\cite{poileUsingComputationalModeling2016}. In particular, we suggest that running preliminary numerical simulations to roughly estimate parameter ranges, and defining informative priors rather than uniform ones, could improve fitting efficiency in future applications; combining multiple experimental conditions may also help reduce the risk of equifinality~\cite{wilsonTenSimpleRules2019}. We see this as a promising research avenue bridging foraging and neuroscience~\cite{grimaForagingEthologicalFramework2025, dennis2026neurobiology, bukwich2025competitive}, with particular relevance for understanding collective decision-making~\cite{gargUtilizingSocialForaging2026a}.\\

Overall, this work provides a unified framework for studying how social information is cognitively represented and integrated, and what consequences it yields on collective decision-making. It offers theoretical insight and practical guidelines for distinguishing between competing hypotheses in collective patch foraging experiments. 

\section*{Acknowledgments}
All authors acknowledge support from the Deutsche Forschungsgemeinschaft (DFG, German Research Foundation) under Germany’s Excellence Strategy, EXC 2117-422037984. A.E.H. is supported by the Human Frontier Science Program (RGP006/2025).
Authors L.B.M. and A.E.H. developed the study, derived the results, and wrote the manuscript.

\section*{Supplementary materials}
\setcounter{figure}{0}
\renewcommand{\thefigure}{S\arabic{figure}}
\captionsetup[figure]{labelformat=sfig, labelsep=period}
\renewcommand{\thetable}{S\arabic{table}}
\captionsetup[table]{labelformat=stable, labelsep=period, labelfont=bf}

\begin{figure}[!h]
    \centering
    \includegraphics[width=0.5\linewidth]{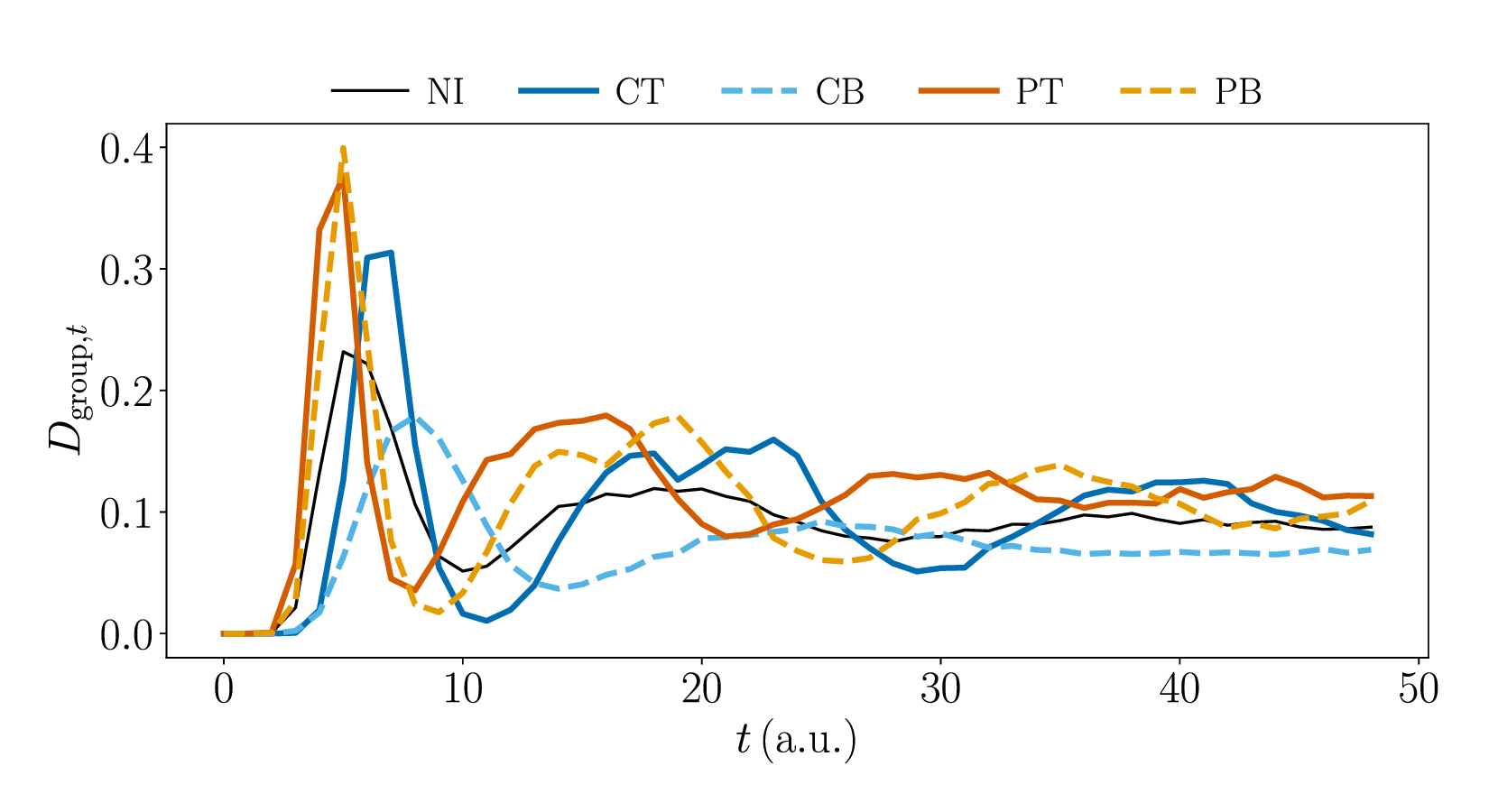}
    \caption{\textbf{Mean intra-group rate of change as a function of time $D_{\mathrm{group},t}$}, for the five socio-cognitive models: NI (black solid line), CT (dark blue solid line), CB (light blue dotted line), PT (dark orange solid line), PB (light orange dotted line).}
    \label{fig:damp}
\end{figure}

\begin{figure}[!h]
    \centering
    \includegraphics[width=\linewidth]{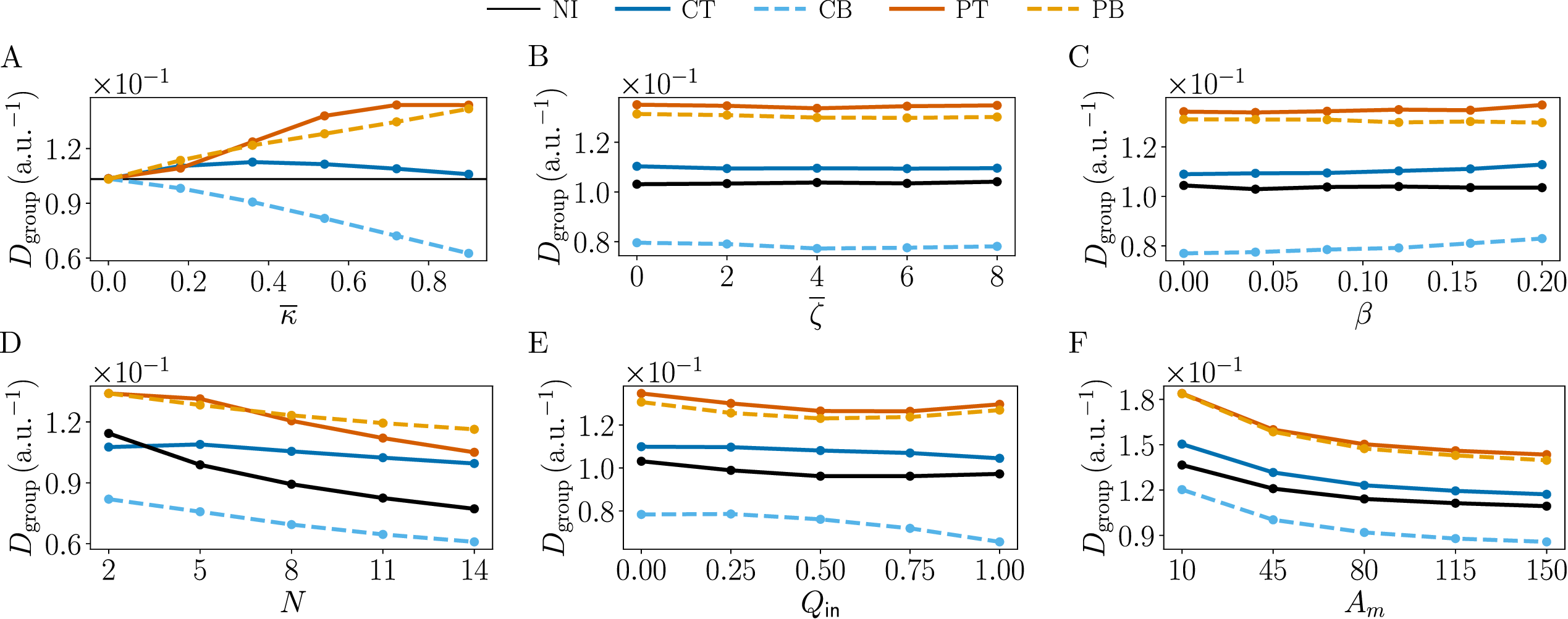}
    \caption{\textbf{Mean intra-group rate of change $D_\mathrm{group}$} for the five socio-cognitive models: NI (black solid line), CT (dark blue solid line), CB (light blue dotted line), PT (dark orange solid line), PB (light orange dotted line), for different parameters: (A) social strength $\overline{\kappa}$, (B) learning parameter $\overline{\zeta}$, (C) heterogeneity factor $\beta$, and experimental conditions: (D) group size $N$, (E) depletion parameter $A_m$, (F) initial accuracy $Q_\text{in}$.}
    \label{fig:D_int}
\end{figure}

\begin{figure}[!h]
    \centering
    \includegraphics[width=0.75\linewidth]{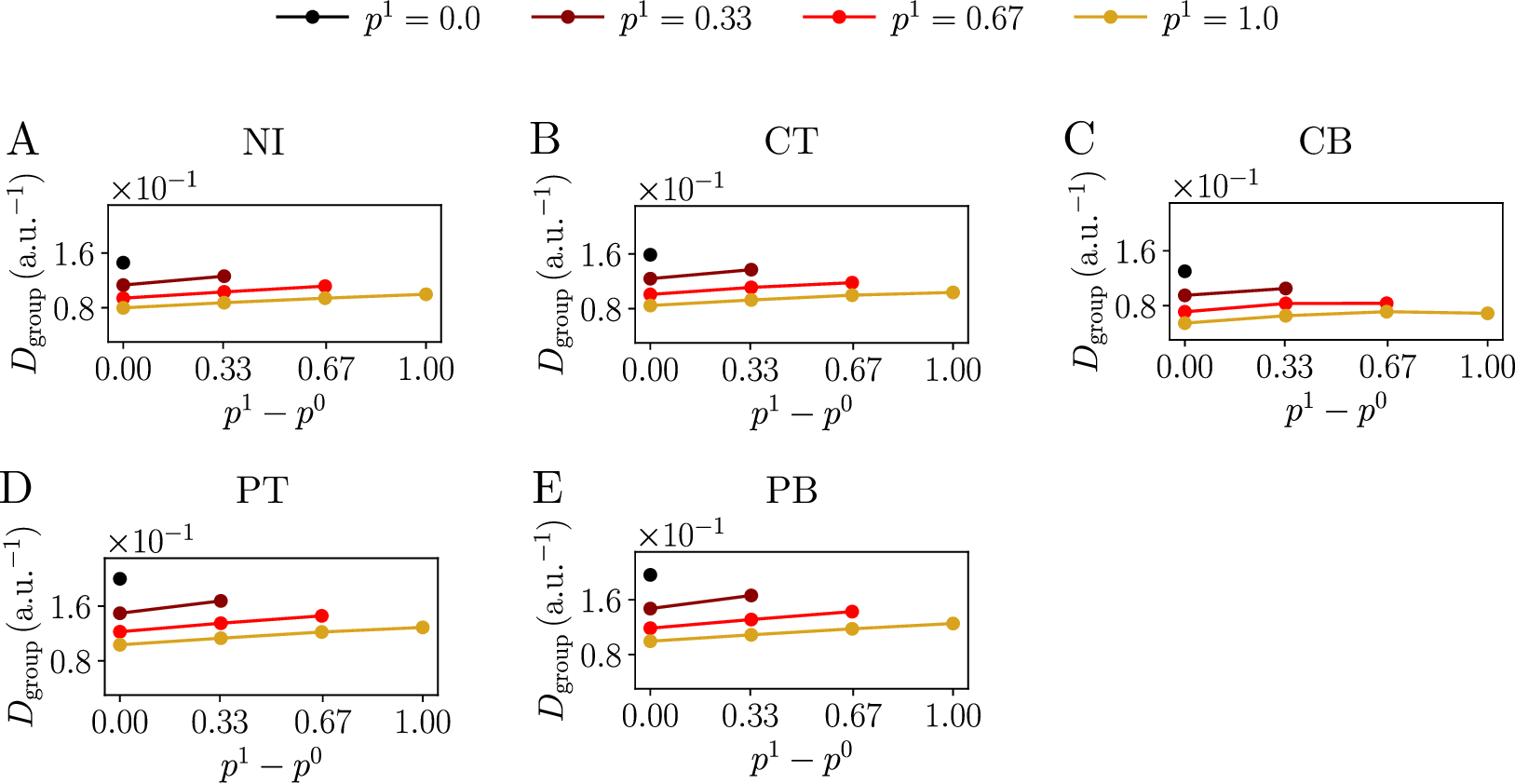}
    \caption{\textbf{Mean intra-group rate of change $D_\mathrm{group}$} for different resource distributions (see color legend on top), for the different cognitive models: (A) NI, (B) CT, (C) CB, (D) PT, (E) PB. }
    \label{fig:D_int_env}
\end{figure}

\begin{table}[htbp]
\centering
\begin{tabular}{lccccc}
\toprule
 & \multicolumn{5}{c}{Fraction of error} \\
\cmidrule(lr){2-6}
Model & $\overline{\theta_0}$ & $\overline{\alpha}$ & $\overline{\zeta}$ & $\overline{\kappa}$ & $\beta$ \\
\midrule
NI & 0.18 & 0.18 & \textbf{0.68} & 0.00 & 0.07 \\
CT & 0.17 & 0.17 & \textbf{0.68}          & 0.05 & 0.08 \\
CB & 0.19 & 0.18 & \textbf{0.67}         & 0.14 & 0.14 \\
PT & 0.05 & 0.05 & \textbf{0.68}          & 0.28 & 0.08 \\
PB & 0.18 & 0.18 & \textbf{0.62}          & 0.02 & 0.09 \\
\bottomrule
\end{tabular}
\caption{\textbf{Fraction of error for each recovered parameter}, by model, for the special case $p^0 = p^1 = 0$. Error is the largest for $\overline{\zeta}$, since in this case the learning term $z_i^k$ is equal to zero, as there is no reward.}
\label{tab:zeta_worse}
\end{table}

\begin{figure}[!h]
    \centering
    \includegraphics[width=0.75\linewidth]{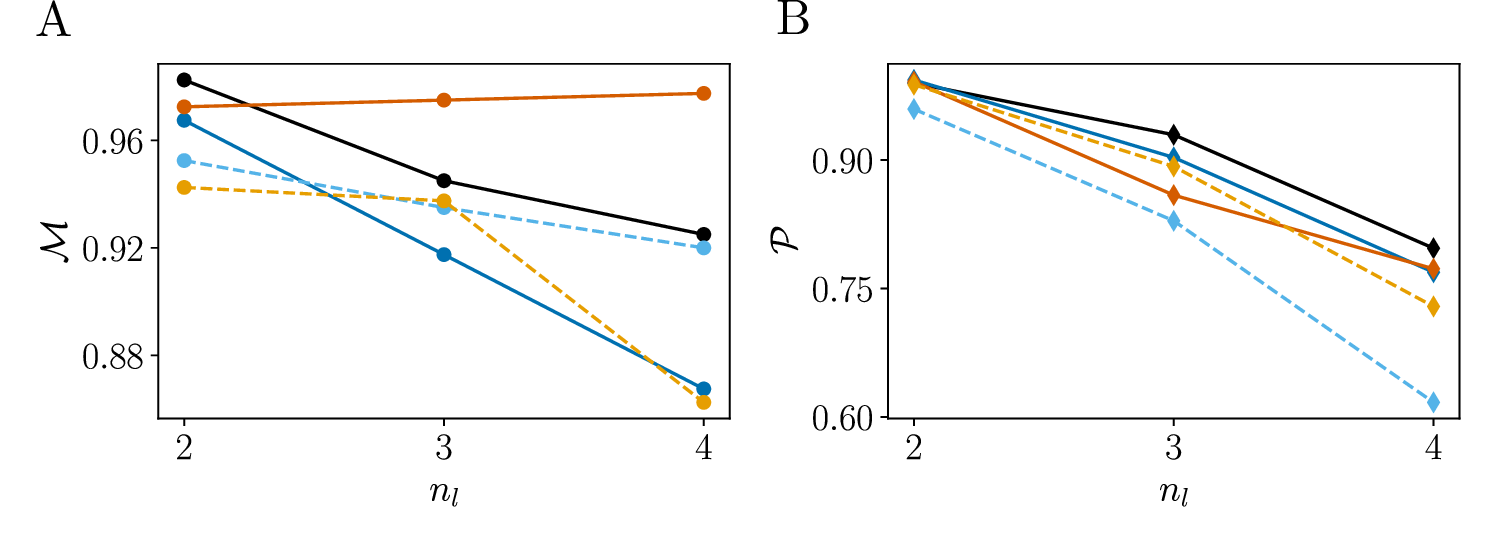}
    \caption{\textbf{Effect of the number of possible values per parameter} ($n_l$) on the proportion of model (A) and parameter (B) correctly predicted between the five socio-cognitive models: NI (black solid line), CT (dark blue solid line), CB (light blue dotted line), PT (dark orange solid line), PB (light orange dotted line), obtained with Bayesian inference.}
    \label{fig:bignl}
\end{figure}

\clearpage
\printbibliography

\end{document}